\documentclass[aps,prc,groupedaddress,twocolumn,preprintnumbers]{revtex4-2}

\usepackage{amsmath,amssymb,bm,mathrsfs}
\usepackage{graphicx,hyperref}
\usepackage{color}

\DeclareMathOperator*{\diag}{\mathrm{diag}}
\DeclareMathOperator*{\Var}{\mathrm{Var}}
\DeclareMathOperator{\PolyLog}{\mathrm{Li}}
\newcommand{\Binomial}[2]{\genfrac(){0pt}{0}{#1}{#2}}
\newcommand{\tBinomial}[2]{\genfrac(){0pt}{1}{#1}{#2}}
\newcommand{\imag}{\mathrm{i}}
\newcommand{\eqdef}{\equiv}
\newcommand{\eqdefr}{\equiv}

\newcommand{\aux}{t}
\newcommand{\Aux}{T}

\begin{document}
\title{Examination of background effects on light-nuclei yield ratio in relativistic heavy-ion collisions}
\author{Shanjin Wu}
\email{shanjinwu2014@pku.edu.cn}
\affiliation{Center for High Energy Physics, Peking University, Beijing 100871, China}
\affiliation{School of Physics and State Key Laboratory of Nuclear Physics and Technology, Peking University, Beijing 100871, China}
\affiliation{Collaborative Innovation Center of Quantum Matter, Beijing 100871, China}
\author{Koichi Murase}
\email{koichi.murase@yukawa.kyoto-u.ac.jp}
\affiliation{Center for High Energy Physics, Peking University, Beijing 100871, China}
\affiliation{Yukawa Institute for Theoretical Physics, Kyoto University, Kyoto 606-8502, Japan}
\author{Shian Tang}
\email{tangshian@pku.edu.cn}
\affiliation{School of Physics and State Key Laboratory of Nuclear Physics and Technology, Peking University, Beijing 100871, China}
\affiliation{Collaborative Innovation Center of Quantum Matter, Beijing 100871, China}
\author{Huichao Song}
\email{huichaosong@pku.edu.cn}
\affiliation{School of Physics and State Key Laboratory of Nuclear Physics and Technology, Peking University, Beijing 100871, China}
\affiliation{Collaborative Innovation Center of Quantum Matter, Beijing 100871, China}
\affiliation{Center for High Energy Physics, Peking University, Beijing 100871, China}
\preprint{YITP-22-57}

\begin{abstract}
The light-nuclei yield ratio is one of the candidates to probe the critical fluctuations of hot QCD matter.
In this paper, we investigate the \textit{background effects}, namely the non-critical effects coming from the non-trivial thermal background, on
the light-nuclei production within the framework of the coalescence model.
Specifically, we analyze the impact of the equilibrium phase-space distribution function of nucleons,
$f(\bm{r},\bm{p})$, on the light-nuclei yield ratio $N_tN_p/N_d^2$, where $N_t$, $N_p$, and $N_d$ denote triton, proton, and deuteron yields.
By considering the characteristic function of the phase-space distribution,
we systematically expand the yield of light nuclei of $A$-constituent nucleons, $N_A$,
in terms of the \textit{phase-space cumulants}, $\langle\bm{r}^n\bm{p}^m\rangle_c$.
We find that the cumulants up to the second-order are canceled out in the generalized ratio $N_p^{B-A} N_B^{A-1}/N_A^{B-1}$.
This means that the dominant background effects including the fireball size,
the kinetic freeze-out temperature, and the coordinate--momentum correlations caused by the radial expansion
play an insignificant role in the yield ratio, which supports the yield ratio as a useful tool for the critical-point search.
We also show several examples of background phase-space distributions for the qualitative illustration.
The higher-order cumulants, which correspond to the non-Gaussian shape of the phase-space profile,
play an important role in the variation of the yield ratio particularly for smaller fireball sizes.
Qualitatively, the spatial structure  of the background decreases the yield ratio, and the azimuthal anisotropy $v_n$ increases it.
The higher order of the azimuthal anisotropy causes a larger effect on the yield ratio.
These results call for the comprehensive future studies of the yield ratio using sophisticated dynamical models.
\end{abstract}
\maketitle
\section{Introduction}\label{sec:intro}
The exploration of the phase structure of Quantum Chromodynamics (QCD) is
one of the main goals of the relativistic heavy-ion experiment
\cite{Stephanov:1998dy,Stephanov:2004wx,Stephanov:2007fk,Asakawa:2015ybt,Aggarwal:2010cw,Luo:2017faz}.
The lattice QCD simulations
\cite{Aoki:2006we,Ding:2015ona,Bazavov:2019lgz,Ratti:2018ksb} have
revealed that the transition from the hadron gas to the quark--gluon
plasma is a crossover at the vanishing baryon chemical potential
($\mu_B\simeq0$).  Meanwhile, calculations based on effective theories of QCD~%
\cite{Klevansky:1992qe,Fukushima:2003fw,Fu:2007xc,Jiang:2013yoa,Roberts:1994dr,Qin:2010nq,Schaefer:2007pw,Pawlowski:2005xe}
conjecture the existence of a critical point along with the
first-order phase transition at finite chemical potential and
temperature. In the vicinity of the phase transition, various variables strongly
fluctuate and could influence the experimental measurements~\cite{Stephanov:1998dy,Stephanov:1999zu,Kitazawa:2012at,Kitazawa:2011wh,Stephanov:2008qz,Stephanov:2011pb,Athanasiou:2010kw,Asakawa:2009aj}.
For instance, the higher-order cumulants of event-by-event multiplicity distribution for the net-proton number measured in the RHIC Beam Energy Scan program exhibit non-monotonic behavior with the varying colliding energy~\cite{STAR:2020tga,STAR:2021iop}, which is claimed as an indication of a potential discovery of the critical point~\cite{Stephanov:2011pb}.

Besides the multiplicity fluctuations of the net-proton number, the yield ratio of the light nuclei
$N_t N_p/N_d^2$~\cite{Sun:2017xrx} obtained by combining the data from NA49
\cite{Anticic:2010mp,Blume:2007kw,Anticic:2016ckv}, STAR
\cite{Adam:2019wnb,Zhang:2019wun}, and ALICE \cite{Adam:2015vda} also
exhibits a non-monotonic behavior as a function of the collision
energy.  Light nuclei are loosely bounded objects with small binding
energy, such as 2.2 MeV for deuteron $d$ and 8.4 MeV for triton $t$.
Their productions are typically described by the coalescence
model~\cite{Schwarzschild:1963zz,Butler:1963pp,Sun:2015jta,Zhu:2015voa}
or the thermal
model~\cite{Andronic:2017pug,Vovchenko:2018fiy,Bellini:2018epz,Xu:2018jff} depending on the
assumption that the light nuclei are formed at the chemical or kinetic freeze-out.
Within the framework of the coalescence model, the light nuclei are formed at the kinetic
freeze-out surface from the nucleons close to each other
in the phase-space because of the small binding energy~%
\cite{Andronic:2017pug,Braun-Munzinger:2018hat}.  It was also pointed
out that the production of light nuclei is related to the relative
neutron density fluctuations~%
\cite{Sun:2017xrx,Sun:2018jhg,Sun:2020pjz,Sun:2020zxy}, and the non-monotonic
behavior may arise from the existence of the critical point~%
\cite{Sun:2017xrx,Sun:2018jhg,Sun:2020pjz,Sun:2020zxy,Shuryak:2019acz,Shuryak:2019ikv,DeMartini:2020hka,DeMartini:2020anq}.
In this direction, intensive studies focus on the light-nuclei
production in relativistic heavy-ion collisions
\cite{Liu:2019nii,Zhao:2018lyf,Zhao:2020irc,Gaebel:2020wid,Deng:2020zxo,Sun:2020uoj,Shao:2020lbq,Kittiratpattana:2020daw}.
However, the interpretation of the observed non-monotonic behavior of
light-nuclei ratio from dynamical models near the phase transition
is still under debate (see e.g., Ref.~\cite{Oliinychenko:2020ply} for a recent review).
One of the complexities comes from the dynamics
near the phase transition, which is expected to have a large impact on the phase-space
distribution at the freeze-out surface.  Different dynamical models near the
critical point and the first-order phase transition have been developed~%
\cite{Nahrgang:2015tva,Asakawa:2015ybt,Yin:2018ejt,Bzdak:2019pkr,Bluhm:2020mpc,Wu:2021xgu,An:2021wof},
but the quantitative comparison with the experimental measurement requires more
sophisticated dynamical modeling, including the equation of
state with a critical point, the proper description of the first-order
phase transition as well as the resonance decays at the later hadronic stage, etc.

Another complexity of using the light-nuclei production to probe the critical fluctuations
is the contamination from the non-critical ones, which we call the \textit{background contributions} hereafter.
In general, the light-nuclei production depends on the coordinate- and momentum-dependence of the emission source~\cite{Monreal:1999mv}, which is
roughly related to the size of the fireball and the scale of the
homogeneity length $l$~\cite{Scheibl:1998tk,Blum:2019suo,Bellini:2020cbj}.  Various factors influence the light-nuclei yields, including the geometry of the
fireball, flow-induced correlations, and resonance decays~\cite{Vovchenko:2020dmv,Oliinychenko:2020znl} besides the critical fluctuations.
Within the framework of the coalescence model, the light-nuclei yields are determined by the phase-space distribution of the
nucleons, $f(\bm{r},\bm{p})$,
which receives the contribution $\delta f_\sigma$ from the critical fluctuations as $f(\bm{r},\bm{p})=f_0(\bm{r},\bm{p})+\delta
f_\sigma(\bm{r},\bm{p})$~\cite{Stephanov:2008qz,Stephanov:2011pb,Jiang:2015hri,Ling:2015yau}.
For the systems not close to the critical point, the magnitude of critical contributions $\delta
f_\sigma(\bm{r},\bm{p})$ is expected to be smaller than the background one $f_0(\bm{r},\bm{p})$ and
make the effects of $\delta f_\sigma$ indistinguishable in the yield.
From another point of view, as the signal, the typical scale of the enhanced critical fluctuations
is characterized by the Kibble--Zurek length $l_{\scriptscriptstyle KZ}$ which is typically smaller than the scale of the homogeneity length
\cite{Wu:2019qfz,Akamatsu:2018vjr} (see also the scale separation of
correlation length and homogeneity length $\xi\ll l$~\cite{Stephanov:2017ghc,Du:2020bxp}).

Within the existing analyses of the coalescence models, the light-nuclei production has been studied
with the assumption of small coordinate--momentum
correlation~\cite{Sun:2017xrx,Sun:2018jhg,Sun:2020pjz},
so that the non-trivial background effects are negligible in the yield.
For example, the coordinate--momentum correlation is typically caused by the collective flow.
When the background flow can be considered uniform within the size of a light nucleus,
we may calculate the yield in the local rest frame where the coordinate--momentum correlation becomes small.
In these studies, the light-nuclei yield was found to be sensitive to the correlation
$\langle \delta \rho_{p,n}(\bm{r})\delta \rho_{p,n}(\bm{r})\rangle$
of the density fluctuation $\delta\rho_{p,n}(\bm{r})\eqdef\rho_{p,n}(\bm{r})-\langle\rho_{p,n}(\bm{r})\rangle$
from the average density of protons or neutrons $\langle\rho_{p,n}(\bm{r})\rangle$,
where $\langle \cdots \rangle\eqdef\frac{1}{V}\int d^3\bm{r}\cdots$
denotes the average over the coordinate space.
The light-nuclei ratio, $N_t N_p/N_d^2$~\cite{Sun:2017xrx}, was found
to be sensitive to the relative density fluctuations normalized by the mean density,
which is in turn sensitive to the fluctuations induced by the phase transition,
and expected to reach a peak at the critical point~\cite{Sun:2020zxy}.
In this existing analysis, a part of the background effect coming from the system size
is effectively eliminated in the ratio,
yet the effect of the inhomogeneous background
at the scale of the light-nuclei size is not considered.

However, in the realistic collision reactions, the background fireballs are not uniform and static
but have finite-size non-trivial shapes and also flows induced by the pressure gradient.
The non-uniform background profiles
of the short scale close to the light-nuclei sizes can be important,
in particular, in smaller collision systems and peripheral collisions.
Even in the central collisions, the background effect can be important
near the edge of the fireball where the gradient of the flow becomes large.
Also, the event-by-event fluctuations coming from the nucleon distribution in the colliding nuclei
may induce the flow fluctuations of shorter scales.
In such situations, the background contributions no longer exactly cancel with each other in the yield ratio;
It is non-trivial how these background configurations affect the yield ratio.
To interpret the experimental data of the yield ratio,
we need to settle these background effects as a baseline.
In this work, we investigate the background effects on the light-nuclei yields and their ratios
by considering how the phase-space distribution plays a role there.
As we focus on the background effects in this paper, we do not consider the critical fluctuations in the present analysis.
We leave the extension with the critical fluctuations in the coming paper, Ref.~\cite{Wu:2022}.
 
Our main strategy in this paper is
(i) to define the \textit{phase-space cumulants}, symbolically denoted as $\langle \bm{r}^n\bm{p}^m\rangle_c$,
using the \textit{characteristic function} $\phi(\bm{t})$ in probability theory,
and (ii) to decompose the phase-space distribution $f_{p,n}(\bm{r},\bm{p})$ in terms of the phase-space cumulants,
and (iii) to study the dependence of the light-nuclei yield ratio on the phase-space cumulants.
Here, $\langle \cdots \rangle = \int d^3\bm{r} d^3\bm{p} f_{p,n}(\bm{r},\bm{p}) \cdots$
is the phase-space average weighted by the phase-space distribution,
and $\langle \cdots\rangle_c$ represents some ``connected'' contributions explained later.
One of our findings is that light-nuclei yields $N_{A}$ and $N_{B}$ with different mass numbers $A$ and $B$ (i.e., the numbers of the constituent nucleons)
share a similar factor carrying the dominant contribution from the second-order cumulants,
which can be exactly canceled in the yield ratios, such as $N_tN_p/N_d^2$ and $N_A^{B-1}N_p^{A-B}/N_B^{A-1}$, assuming the same light-nuclei radii.
The resulting yield ratios are insensitive to the size of the fireball
and to the coordinate--momentum correlation caused by the collective expansion,
up to the second order in the phase-space cumulants.
On the other hand, the higher-order phase-space cumulants,
which correspond to the non-Gaussian nature of the phase-space distribution,
do not cancel with each other in the yield ratio.
We conclude that the non-Gaussian shape in the phase-space distribution plays an important role
in the interpretation of the observed non-monotonic behavior of the light-nuclei yield ratio~\cite{Zhang:2019wun,Liu:2019nii}.

This paper is organized as follows:
After briefly introducing the light-nuclei production in the coalescence model in Sec.~\ref{sec:coalescence},
we introduce the standardized Jacobi coordinates for the Wigner function in Sec.~\ref{sec:phasevariable} for later convenience.
In Sec.~\ref{sec:formalism}, we derive the light-nuclei production in terms of the phase-space cumulants and discuss the yield ratio.
In Sec.~\ref{sec:example}, we see various examples of the background phase-space distribution
including the Gaussian (Sec.~\ref{sec:example.gauss}),
Woods--Saxon (Sec.~\ref{sec:example.ws}),
and the double-Gaussian (Sec.~\ref{sec:example.double}) spatial profiles
as well as the Gaussian profile with the radial expansions (Sec.~\ref{sec:example.radial})
and anisotropic flows (Secs.~\ref{sec:example.elliptic} and \ref{sec:example.blastwave}).
Finally, we summarize and conclude this paper in Sec.~\ref{conclusion}.
In this paper, the natural unit system $k_B = \hbar = c = 1$ is adopted.

\section{Light-nuclei production in the coalescence model}\label{sec:coalescence}
In the coalescence model~\cite{Schwarzschild:1963zz,Butler:1963pp,Sun:2015jta,Zhu:2015voa},
the production of the light-nuclei is determined by the overlap of the
phase-space distribution functions of nucleons $f_{i}(\bm{r},\bm{p})$  with the
Wigner function
$W_A(\{\bm{r}_i,\bm{p}_i\}^A_{i=1})$ of the nucleus at the
freeze-out:
\begin{align}\label{eq:coale-1}
    N_A=g_A \int \biggl[\prod^A_i d^3\bm{r}_i d^3\bm{p}_if_i(\bm{r}_i,\bm{p}_i)\biggr] W_A(\{\bm{r}_i,\bm{p}_i\}^A_{i=1}),
\end{align}
where $g_A=(2s+1)/2^A$ is
the statistical factor with spin $s$ for nuclei of the mass number $A$~\cite{Sato:1981ez}.
It should be noted here that
the light-nuclei yield measured in experimentals
corresponds to the number of nuclei with the finite acceptance of e.g. rapidity.
To calculate the quantitative yields that can be directly compared with the data,
we need to introduce constraints of the total momentum in the integration domain of momenta $\int d^{3A}\bm{p}$.
Currently, we focus on the qualitative nature of the background effects.
We thus integrate over the whole range of the momentum for simplicity
assuming that the qualitative behavior is not largely affected by the finite acceptance.

The Wigner function $W_A(\{\bm{r}_i,\bm{p}_i\}^A_{i=1})$ is obtained by the Wigner
transform of the wave function, which is assumed to be the
spherical harmonic-oscillator wave function~\cite{Scheibl:1998tk,Aerts:1983hy}:
\begin{align}
W_A(\{\bm{r}_i,\bm{p}_i\}^A_{i=1})=8^{A-1}\exp\biggl[-\sum^{A-1}_{i=1}\biggl(\frac{\bm{R}^2_i}{\sigma_{(i)}^2}+\sigma_{(i)}^2\bm{P}^2_i\biggr)\biggr], \label{eq:wigner-1}
\end{align}
where the relative coordinates and momentum, $\bm{R}_i$ and $\bm{P}_i$,
are defined in the Jacobi coordinates with the mass of $i$-th nucleon being $m_i$~\cite{Mattiello:1996gq,Chen:2003ava,Sun:2015jta}.
In this work, we assume $m_1=\cdots=m_A=m$ since we focus on the light nuclei
which consist of protons and neutrons with almost the same masses.
In this case, the Jacobi coordinates are simply written as
\begin{align}\label{eq:relative-RP}
    \begin{pmatrix}
      \bm{R}_i\\
      \bm{P}_i
    \end{pmatrix}&\eqdef\sqrt{\frac{i}{i+1}}\biggl[
    \frac 1i
    \sum^i_{j=1}
    \begin{pmatrix}
    \bm{r}_j\\
    \bm{p}_j
    \end{pmatrix}
    -
    \begin{pmatrix}
      \bm{r}_{i+1}\\
      \bm{p}_{i+1}
    \end{pmatrix}
    \biggr], \\
    &\qquad (i=1,\dots,A-1),\nonumber\\
    \begin{pmatrix}
      \bm{R}_A\\
      \bm{P}_A
    \end{pmatrix}&\eqdef
    \frac{1}{\sqrt{A}}\sum^A_{j=1}
    \begin{pmatrix}
    \bm{r}_j\\
    \bm{p}_j
    \end{pmatrix},
\end{align}
The width parameter for the Wigner function is $\sigma_{(i)}=(m_i
\omega)^{-1/2} = (m\omega)^{-1/2}$, where the harmonic-oscillator frequency $\omega$ is
related to the root-mean-square radius $r_A$ of the nucleus $A$~\cite{Sun:2015jta} as
\begin{align}
  \sigma_{(1)}^2 = \dots = \sigma_{(A)}^2 = \frac{2A}{3(A-1)}r_A^2 \eqdefr \sigma_A^2.
\end{align}
For example, the values of the width parameters for deuteron, triton, and helium-4
are $\sigma_d=2.26$, $\sigma_t=1.59$, and $\sigma_{^4\mathrm{He}}=1.37$ fm~\cite{Ropke:2008qk}, respectively. In this work, we assume
the same light-nuclei radii $\sigma_d\simeq \sigma_t \simeq
\sigma_{^4\mathrm{He}}$ or $\sigma_A \simeq \sigma_B$ (for the nuclei $A$ and $B$) unless otherwise specified.

In relativistic heavy-ion collisions, the phase-space distributions $f_{p,n}(\bm{r},\bm{p})$ fluctuate from event to event.
In this study, we consider a single phase-space distribution
which can be, for example, identified as the average phase-space distribution over collision events:
\begin{align}\label{eq:averaged-f}
    \bar{f}_{p,n}(\bm{r},\bm{p})\eqdef \langle f_{p,n}(\bm{r},\bm{p})\rangle_\text{ev}.
\end{align}
This means that we miss possible effects of the event-by-event fluctuations in the present analysis.
In fact, the effect of the event-by-event fluctuations on the phase-space distribution,
and in turn on the two-point correlation, has been addressed in Hanbury-Brown--Twiss (HBT)
femtoscopy~\cite{Plumberg:2015mxa,Plumberg:2013nga}, especially for
the third-order oscillations of the HBT radii which is found to be sensitive to the
initial state fluctuations~\cite{Plumberg:2013nga}.
Thus we anticipate that the event-by-event fluctuations can be important also in the light-nuclei yields
yet leave it for future study as we focus on the average part of the background effects in this work.
For simplicity of the notation,
we use $f_{p,n}(\bm{r},\bm{p})$ to represent the average phase-space
distribution~\eqref{eq:averaged-f} hereafter.
For the present study, we do not consider the isospin asymmetry, and thus
$f(\bm{r},\bm{p}) \eqdef f_{p}(\bm{r},\bm{p}) = f_{n}(\bm{r},\bm{p})$.

\subsection{Example for Gaussian phase-space distribution}\label{sec:coalescence.gauss}
Here, we demonstrate an example of the light-nuclei yields
under a simple Gaussian phase-space distribution and observe the result.
Following Refs.~\cite{Sun:2018mqq, Sun:2020pjz, ExHIC:2011say, Mrowczynski:2020ugu},
we assume the following form of the phase-space distribution with Gaussian forms in both momentum and position space:
\begin{align}
    \label{eq:dist-gauss}
    f(\bm{r},\bm{p}) &= \frac{\rho_0}{(2\pi mT)^{3/2}} \exp\Bigl(-\frac{\bm{r}^2}{2R_s^2}\Bigr)\exp\Bigl(- \frac{\bm{p}^2}{2mT}\Bigr),
\end{align}
where $T$ is the freeze-out temperature,
$R_s$ is the typical fireball size in the coordinate space,
and the core density $\rho_0 = N_p/(2\pi R_s^2)^{3/2}$ is fixed as a normalization parameter to satisfy $N_p=\int d^3\bm{r}d^3\bm{p}f(\bm{r},\bm{p})$.

With such a phase-space density, the light-nuclei yield~\eqref{eq:coale-1} is reduced to simple Gaussian integrations (see Appendix~\ref{app:simple-gauss}) so that the result reads
\begin{align}
  N_d &= g_d N_p^2 \Bigl[\Bigl(R_s^2 + \frac{\sigma_d^2}2\Bigr)\Bigl(mT + \frac1{2\sigma_d^2}\Bigr)\Bigr]^{-\frac32}, \\
  N_t &= g_t N_p^3 \Bigl[\Bigl(R_s^2 + \frac{\sigma_t^2}2\Bigr)\Bigl(mT + \frac1{2\sigma_t^2}\Bigr)\Bigr]^{-3}.
  \label{eq:gaussian-yields}
\end{align}
Here, we find that these light-nuclei yields exactly cancel with each other in the yield ratio
if the effects of the different nuclei-nuclei sizes $\sigma_d$ and $\sigma_t$
are assumed to be negligible compared to the terms $R_s^2$ and $mT$.
As a result, the yield ratio becomes a constant $N_t N_p/N_d^2 = g_t/g_d^2 = 4/9$
solely fixed by the spin of deuterons and tritons.
This is a remarkable result that
the yield ratio is actually insensitive to the fireball size
and the momentum distribution
as far as the phase-space distribution can be approximated by Eq.~\eqref{eq:dist-gauss}.

However, the realistic phase-space distribution in relativistic heavy-ion collisions
cannot be correctly captured by the form~\eqref{eq:dist-gauss}.
The most phenomenologically-important missing ingredient is the effect of the collective flow.
Eq.~\eqref{eq:dist-gauss} has the factorized form
for the coordinate profile and the momentum distribution,
i.e., coordinates and momentum are independent of each other.
In reality, however, hydrodynamic expansions and anisotropic flow induce strong coordinate--momentum correlations.
Another missing ingredient is the detailed structure of the phase-space profile of the fireballs,
which cannot be fully captured by a simple Gaussian form.
Existing works~\cite{Zhao:2018lyf,Zhao:2020irc,Hillmann:2021zgj}
have already numerically considered specific phase-space distributions from hydrodynamic models
to calculate the light-nuclei production in relativistic heavy-ion collisions,
but we rather analytically approach the light-nuclei yields for the general phase-space distribution
to establish a physical picture of the relation between the background profile and the yield ratio.
We will address these non-trivial ingredients for the yield ratio in the next sections.
We also check the effect of different light-nuclei sizes $\sigma_t \neq \sigma_d$ later.

\section{Standardized Jacobi coordinates for the Wigner function}\label{sec:phasevariable}
As explained in Sec.~\ref{sec:coalescence}, the probability for the
production of light nuclei is determined by the relative distance in
the phase-space, which is characterized by the Wigner function.
To take advantage of this property,
we here introduce new variables, reorganize the notation,
and observe interesting consequences
which play an important role in later sections.

Also, to consider the coordinate--momentum correlation, it is convenient to introduce a uniform notation for $\bm{r}_i$ and $\bm{p}_i$.
We here rescale the phase-space coordinates {$(\bm{r},\bm{p}) \in \mathbb{R}^6$ and redefine it as $\bm{z} \in \mathbb{R}^6$}:
\begin{align}
    &\begin{pmatrix}
    \bm{z}_1,&\bm{z}_2,&\cdots,&\bm{z}_A
    \end{pmatrix}^\mathrm{T}\nonumber\\
    &\eqdef\sqrt{2}
    \begin{pmatrix}
      \frac{\bm{r}_1}{\sigma_A},& \sigma_A \bm{p}_1,& \frac{\bm{r}_2}{\sigma_A},&\sigma_A \bm{p}_2,& \cdots,& \frac{\bm{r}_A}{\sigma_A},&\sigma_A \bm{p}_A
    \end{pmatrix}^\mathrm{T}. \label{eq:standard-phasespace}
\end{align}
The corresponding relative distance $\bm{Z}$ is defined as
\begin{align}
    &\begin{pmatrix}
    \bm{Z}_1,&\bm{Z}_2,&\cdots,&\bm{Z}_A
    \end{pmatrix}^\mathrm{T}\nonumber\\
    &\eqdef\sqrt{2}
    \begin{pmatrix}
      \frac{\bm{R}_1}{\sigma_A},& \sigma_A \bm{P}_1,& \frac{\bm{R}_2}{\sigma_A},&\sigma_A \bm{P}_2,& \cdots,& \frac{\bm{R}_A}{\sigma_A},&\sigma_A \bm{P}_A
    \end{pmatrix}^\mathrm{T},
\end{align}
with which the Wigner function \eqref{eq:wigner-1} is re-expressed as
\begin{align}\label{eq:wigner-2}
    W_A(\{\bm{r}_i,\bm{p}_i\}^A_{i=1}) =8^{A-1}\exp\biggl[-\frac{1}{2}\sum^{A-1}_{i=1}\bm{Z}_i^2\biggr].
\end{align}
Here, $\bm{Z}_i$ are identified to be the \textit{standardized variables} of the Gaussian form in the Wigner function,
i.e., the Gaussian widths of $\bm{Z}_i$ are standardized to unity.
In this notation, the transformation~\eqref{eq:relative-RP} between $(\bm{R}_i,\bm{P}_i)$ and $(\bm{r}_i,\bm{p}_i)$ takes a simpler form:
\begin{align}\label{eq:zZ}
    \bm{Z}_i=\sqrt{\frac{i}{i+1}}\biggl(\frac{1}{i}\sum^i_{j=1}\bm{z}_j-\bm{z}_{i+1}\biggr),\qquad \bm{Z}_A=\frac{1}{\sqrt{A}}\sum^A_{i=1}\bm{z}_i.
\end{align}
This linear transformation $O$: $\bm{Z}_i = \sum_{j=1}^A O_{ij} \bm{z}_j$ between $\bm{Z}_i$ and $\bm{z}_i$ is an orthogonal transformation,
i.e., $O^\mathrm{T} O = 1$, so that the ``norm'' is preserved:
\begin{align}\label{eq:zZ2}
    \sum^{A-1}_{i=1}\bm{Z}^2_i + \bm{Z}_A^2 = \sum^A_{i=1}\bm{Z}^2_i = \sum^A_{i=1}\bm{z}^2_i.
\end{align}

This means that the cross terms $\bm{z}_i\cdot\bm{z}_j$,
which originally appeared in the Wigner function~\eqref{eq:wigner-2} when $\bm{Z}_i$ are expanded in terms of $\bm{z}_i$,
exactly cancel with those that appear from $\bm{Z}_A^2$.
This property is expected from the original idea of the Wigner function
written in the relative coordinates~\eqref{eq:relative-RP}.
The property can be ultimately attributed to the translational invariance of
the interaction between the constituent nucleons against the center-of-mass motion,
which enables us to write the nuclear interaction
using only the relative coordinates between the nucleons, $\bm{Z}_i$, ($i=1,\dots,A-1$).
For this reason, for the harmonic-oscillator Wigner function,
the Jacobi coordinates $\bm{Z}_i$ are useful to separate the center-of-mass motion
from the total squared distance $\sum^A_{i=1}\bm{z}^2_i$:
\begin{align}
  \sum^A_{i=1}(\bm{z}_i-\bm{Z}_\mathrm{cm})^2
  &= \sum^A_{i=1}\bm{z}_i^2 -A\bm{Z}_\mathrm{cm}^2 \nonumber \\
  &= \sum^A_{i=1}\bm{Z}_i^2 - \bm{Z}_A^2
  = \sum^{A-1}_{i=1}\bm{Z}_i^2,
\end{align}
where $\bm{Z}_\mathrm{cm}\eqdef(1/A) \sum_{i=1}^A\bm{z}_i = \bm{Z}_A/\sqrt{A}$ is the center-of-mass coordinates, which disappears in Eq.~\eqref{eq:wigner-2}.
The transformation from the phase-space coordinates
$\bm{z}_i$ to the Jacobi coordinates $\bm{Z}_i$ {leads to} the expression in which the center-of-mass
motion of the nucleus is explicitly separated from the internal motion of the
nucleons.

Finally, the Jacobian determinants for the transformation from $(\bm{r}_i,\bm{p}_i)$ to $\bm{z}_i$ and $\bm{Z}_i$ are given by
$8^A \prod^A_{i=1}d^3\bm{r}_i d^3\bm{p}_i
= 8^A \prod^A_{i=1}d^3\bm{R}_i d^3\bm{P}_i
= \prod^A_{i=1}d^6\bm{z}_{i} = \prod^A_{i=1}d^6\bm{Z}_{i}$.

\section{Phase-space distribution and light-nuclei production}\label{sec:formalism}

We have observed in Sec.~\ref{sec:coalescence.gauss} that
the Gaussian phase-space profile $f(\bm{r},\bm{p})$ produces a constant yield ratio.
We anticipate that the deviation of the phase-space distribution from the Gaussian profile
would give corrections to the yield ratio.
Here, the goals are to systematically decompose the expression of the light-nuclei yield into the Gaussian part and the deviation
and to investigate the impact of the deviation on the yield ratio.
One of such methods can be formulated by considering the cumulants of the phase-space distribution.
We here introduce a new formalism of writing the light-nuclei yield down in terms of various orders of
cumulants of the phase-space variables $(\bm{r},\bm{p})$.

\subsection{Characteristic function and cumulants}\label{sec:formalism.cumulants}
In this subsection, we summarize the basics of the characteristic function and the cumulants.
The readers who are familiar with them can safely skip this subsection.
In probability theory, an arbitrary distribution, $\rho(\bm{z})$ (where $\bm{z}\in\mathbb{R}^m$), can be specified by
the corresponding \textit{characteristic function}, which is nothing
but the Fourier transform of the probability density function:
\begin{align}
  \phi(\bm\aux)&=\langle e^{i\bm{z}\cdot\bm\aux}\rangle = \int d^m\bm{z} \rho(\bm{z}) e^{i\bm{z}\cdot\bm\aux}.
  \label{eq:cf1}
\end{align}
When the distribution has bounded moments,
the characteristic function can be used as the moment-generating function,
where the parameter $\bm\aux$ plays an analogous role of an ``external field'' coupled to the random variable $\bm{z}$.

The logarithm of the characteristic function is known as the \textit{second cumulant-generating function}:
\begin{align}
  \ln \phi (\bm\aux) = \sum_{\bm{\alpha} \in \mathbb{N}_0^m} \frac{\mathcal C_{\bm{\alpha}}}{\bm{\alpha}!} (\imag\bm\aux)^{\bm{\alpha}}.
\end{align}
For simplicity, we here introduced the multi-index notation
$\bm{\alpha}=(\alpha_1,\alpha_2,\cdots,\alpha_m)\in
\mathbb{N}_0^m=\{0,1,2,\ldots\}^m$ for the $m$ components
of $\bm{z}^{\bm{\alpha}}$ or $\bm\aux^{\bm{\alpha}}$,
where $\bm\aux^{\bm{\alpha}}$ is a shorthand
for $\bm\aux^{\bm{\alpha}} \eqdef \aux_{1}^{\alpha_1}\cdot \aux_{2}^{\alpha_2}\cdots \aux_{m}^{\alpha_m}$.
Likewise, the symbols $|\bm{\alpha}| \eqdef \alpha_1+\cdots+\alpha_m$
and $\bm{\alpha}! \eqdef \alpha_1!\cdots \alpha_m!$ denote
the degree of the index and the factorials, respectively.
The coefficient $\mathcal{C}_{\bm{\alpha}} \eqdef (-\imag\partial_{\bm\aux})^{\bm{\alpha}}\ln\phi(\bm\aux)|_{\bm\aux=0}$
can be identified as the cumulant of $\bm{z}$ of the order $|\bm{\alpha}|$,
$\mathcal{C}_{\bm{\alpha}}\eqdef\langle \bm{z}^{\bm{\alpha}}\rangle_c$,
which is defined as
\begin{align}\label{def:cumulants}
  \langle \bm{z}^{\bm{\alpha}}\rangle_c
  &=\sum_{\beta}(|\beta|-1)! (-1)^{|\beta|-1} \prod_{B\in\beta} \int d^m\bm{z} \rho(\bm{z})\prod_{i\in B} X_i.
\end{align}
In the above summation and the products, $\beta$ runs through all the possible
partitions of $\{1,2,\dots,|\alpha|\}$, $B$ runs through the list of
blocks in the partition $\beta$, and $|\beta|$ is the number of blocks
in the partition.  In the case of $\bm\alpha=\bf0$,
there is no partition $\beta$ for the empty set so that $\mathcal C_{\bm 0} = \langle \bm{z}^{\bm0}\rangle_c = 0$.
The variables $X_i$ denote the multiset of $\bm{z}$ weighted by the \textit{multiplicity} $\bm{\alpha}$, i.e.,
\begin{align}
(X_1,X_2,\dots,X_{|\bm{\alpha}|}) &= (\overbrace{z_1,\dots,z_1}^{\text{$\alpha_1$ times}},
\cdots,\overbrace{z_m,\dots,z_m}^{\text{$\alpha_m$ times}}).
\end{align}
The second cumulant-generating function, $\ln \phi(\bm{t})$, can be compared with the generating functional of the connected diagrams, $W[J]$, in the field theory,
and the cumulants can be regarded as the connected contributions of the moments which cannot be explained by the lower-order cumulants.

A remarkable fact is that,
given that the cumulants of all the orders are well-defined and known
and that the series \eqref{eq:cf1} is convergent in a neighborhood of $\bm{t}=0$
(see Refs.~\cite{Akhiezer:1965,Devroye:1986} for the necessary and sufficient condition called Carleman's condition),
the original distribution $\rho(\bm{z})$ can be uniquely reconstructed by the set of cumulants
through the inverse Fourier transform of the characteristic function:
\begin{align}
  \rho(\bm{z}) &= \int\frac{d^m\bm\aux}{(2\pi)^m} e^{-\imag\bm\aux\cdot\bm{z}} \exp \biggl[\sum_{\bm{\alpha}\in\mathbb{N}_0^m} \frac{\mathcal{C}_{\bm{\alpha}}}{\bm{\alpha} !}(\imag \bm\aux_i)^{\bm{\alpha}}\biggr],
\end{align}
which justifies the discussion using an arbitrary distribution just with moments and cumulants.

As an example, let us consider the general Gaussian distribution of $m$ variables:
\begin{align}
  \rho(\bm{z}) = \frac1{\sqrt{(2\pi)^m \det G}}
    \exp\Bigl[-\frac12(\bm{z}-\bm{z}_c)^\mathrm{T} G^{-1}(\bm{z}-\bm{z}_c)\Bigr],
\end{align}
with $\bm{z}_c \eqdef \langle\bm{z}\rangle_c = \langle\bm{z}\rangle$
and $G \eqdef \langle \bm{z}\bm{z}^\mathrm{T}\rangle_c = \mathrm{Cov}_{\bm{zz}}$
being the mean and the covariance matrix of $\bm{z}$, respectively.
It is ready to calculate the all-order cumulants
using the definition of the characteristic function~\eqref{eq:cf1}:
\begin{align}
  \ln\phi(\bm\aux) = i\bm{z}_c\cdot\bm\aux - \frac12\bm\aux^\mathrm{T} G \bm\aux,
\end{align}
where we find that the cumulant-generating function only has two terms up to the second-order of $\bm{t}$.
This means that all the higher-order cumulants
$\{\mathcal{C}_{\bm\alpha}\}_{|\bm{\alpha}|\ge3}$ vanish for the Gaussian distribution.
In other words, all the higher-order cumulants can be regarded as
the parameters of the non-Gaussian distortion of the distribution
from the underlying Gaussian distribution.
In this way, the distribution can be decomposed into
the underlying Gaussian distribution specified by the cumulants up to the second order
and the non-Gaussian components specified by the high-order cumulants.
This is useful for the present purpose of
studying the effect of the phase-space distribution deformation
from the Gaussian distribution as discussed in Sec.~\ref{sec:coalescence.gauss}.

\subsection{Formalism}\label{sec:formalism.yield}

Now, we can express the phase-space distribution of the nucleons
in Eq.~\eqref{eq:coale-1} in terms of the various orders of cumulants.
We define the distribution $f(\bm{z}_i)$ as a function of $\bm{z}_i$:
\begin{align}\label{def:distribution-in-z}
  f(\bm{z}_i) d^6\bm{z}_i \eqdef f(\bm{r}_i,\bm{p}_i) d^3\bm{r}_i d^3\bm{p}_i,
\end{align}
so that $f(\bm{z}_i) = |\partial (\bm{r}_i,\bm{p}_i)/\partial (\bm{z}_i)|
  f(\bm{r}_i,\bm{p}_i) = 8^{-1}f(\bm{r}_i,\bm{p}_i)$,
where $|\partial (\bm{r}_i,\bm{p}_i)/\partial (\bm{z}_i)|$ is the Jacobian determinant.
The distribution is decomposed as
\begin{align}\label{eq:dist.cumulants}
  \frac{f(\bm{z}_i)}{N_p}
    &= \rho(\bm{z}_i) = \int \frac{d^6\bm\aux_i}{(2\pi)^6} e^{-\imag\bm\aux_i\cdot \bm{z}_i}
    \exp \biggl[\sum_{\bm{\alpha}\in\mathbb{N}_0^6} \frac{\mathcal{C}_{\bm{\alpha}}}{\bm{\alpha} !}(\imag \bm\aux_i)^{\bm{\alpha}}\biggr],
\end{align}
where the normalization $N_p = N_n = \int d^6\bm{z} f(\bm{z})$ is given by the number of each nucleon species,
$\bm{z}_i\in \mathbb{R}^6$ is the redefined position of $i$-th nucleon in 6-dimensional phase-space in Eq.~\eqref{eq:standard-phasespace},
and $\bm\aux_i$ is the corresponding Fourier conjugate with $i=\{1,2,\dots,A\}$ representing the index of nucleons inside the light nuclei.

In this context, $\mathcal{C}_{\bm{\alpha}}\equiv\langle \bm{z}^{\bm{\alpha}}\rangle_c$
is the cumulant of the phase-space variables $\bm{z}_i=\sqrt{2}(\bm{r}_i/\sigma_A,\sigma_A\bm{p}_i)$ of the order $|\bm{\alpha}|$,
where $\langle\cdots\rangle = (1/N_p)\int d^6\bm{z} f(\bm{z}) \cdots$ denotes
the averaging over the phase-space under a single phase-space distribution $f(\bm{r},\bm{p})$.
It should be noted that these are completely different types of cumulants
than the typical event-by-event cumulants of, e.g., the anisotropic flows
or the multiplicity distribution for the net-proton number
where $\langle\cdots\rangle$ denotes the event-by-event average.
In this paper, we call these cumulants of the phase-space variables
the ``\textit{phase-space cumulants}'' hereafter.

To calculate the yield, we substitute
Eqs.~\eqref{eq:wigner-2}, \eqref{def:distribution-in-z} and \eqref{eq:dist.cumulants} for Eq.~\eqref{eq:coale-1},
perform the integration by $\prod_{i=1}^Ad^6\bm{z}_i = \prod_{i=1}^Ad^6\bm{Z}_i$,
and obtain the following formula:
\begin{align}
  N_A &= g_A 8^{A-1} N_p^A
    \int \biggl[\prod_{i=1}^A \frac{d^6\bm\Aux_i}{(2\pi)^3}\biggr]
    (2\pi)^3 \delta^{(6)}(\bm\Aux_A) \nonumber\\
  &\quad\times
    \exp\biggl[-\frac12\sum_{i=1}^{A-1}\bm\Aux_i^2
      + \sum_{i=1}^A\sum_{\bm{\alpha}\in\mathbb{N}_0^6} \frac{\mathcal C_{\bm\alpha}}{\bm\alpha!}(\imag\bm\aux_i)^{\bm\alpha}\biggr].
\end{align}
In the above summation with respect to $\bm\alpha$,
the term of $|\bm\alpha|=0$ vanishes due to $\mathcal{C}_{\bm0}=0$.
The sum of the first-order terms of $|\bm\alpha|=1$, which is proportional to $\bm\Aux_A$,
also vanishes due to the delta function $\delta^{(6)}(\bm\Aux_A)$.
This is a consequence of the translational symmetry of the Wigner function for the center-of-mass motion.
The second-order terms can be transformed by the orthogonal transformation
$\bm\Aux_i = O_{ij} \bm\aux_j$ as
\begin{align}
  \sum_{i=1}^A\sum_{|\bm{\alpha}|=2}
    \frac{\mathcal C_{\bm\alpha}}{\bm\alpha!}(\imag\bm\aux_i)^{\bm\alpha}
  &= -\frac12 \sum_{i=1}^A\sum_{\beta,\gamma=1}^6
    \mathcal{C}_{2(\beta,\gamma)} \aux_{i,\beta} \aux_{i,\gamma} \nonumber \\
  &= -\frac12 \sum_{i=1}^A\sum_{\beta,\gamma=1}^6
    \mathcal{C}_{2(\beta,\gamma)} \Aux_{i,\beta} \Aux_{i,\gamma},
\end{align}
where $\mathcal C_{2(\beta,\gamma)} \eqdef \langle z_\beta z_\gamma\rangle_c$ is the second-order cumulants.
We used the property $O^\mathrm{T} O = 1$ to obtain the second line.
Finally, the light-nuclei yield can be written in the form of the Gaussian integration:
\begin{align}\label{eq:NA4}
  N_A &= g_A 8^{A-1} N_p^A
    \int \biggl[\prod_{i=1}^{A-1} \frac{d^6\bm\Aux_i}{(2\pi)^3}\biggr]
    \sum_{m=0}^\infty \frac{[\mathcal H(\{\bm{T}_i\}_{i=1}^{A-1})]^m}{m!} \nonumber\\
  &\quad\times
    \exp\biggl[
      -\frac12 \sum_{i=1}^{A-1}\sum_{\beta,\gamma=1}^6
      (\mathcal{C}_{2(\beta,\gamma)} + \delta_{\beta,\gamma})
      \Aux_{i,\beta} \Aux_{i,\gamma}\biggr],
\end{align}
where $\mathcal H$ contains the higher-order terms
written as a polynomial of $\{\bm\Aux_i\}_{i=1}^{A-1}$:
\begin{align}
  \mathcal H(\{\bm{T}_i\}_{i=1}^{A-1})
    &\eqdef
     \sum_{k=3}^\infty \mathcal H_k(\{\bm{T}_i\}_{i=1}^{A-1}) \nonumber \\
    &\eqdef
      \sum_{k=3}^\infty \sum_{i=1}^A \sum_{|\bm{\alpha}| = k}
      \frac{\mathcal C_{\bm\alpha}}{\bm\alpha!}[\imag (O^\mathrm{T} \bm\Aux)_i]^{\bm\alpha}
      \biggr|_{\bm{T}_A = 0}.
\end{align}

When the distribution is sufficiently close to the Gaussian distribution
so that the higher-order cumulants are sufficiently small,
we may perturbatively calculate the yield
using Wick's theorem for the Gaussian integration,
where $(\mathcal{C}_{2(\beta,\gamma)} + \delta_{\beta,\gamma})^{-1}$ and $\mathcal H_k$
are identified to be the analogs of the ``propagator'' and the ``$k$-point coupling'', respectively.
It should be noted that the cumulant expansions can be asymptotic expansions for the general distribution,
so the perturbation might need to be explicitly truncated to a finite order for the optimal approximation.
The lowest order with respect to $m$ in Eq.~\eqref{eq:NA4} reads
\begin{align}
  N_A^{(0)} &\eqdef
  g_A N_p^A 8^{A-1}[\det(\mathcal C_2 + \mathcal I_6)]^{-(A-1)/2},
  \label{eq:NA-lowest}
\end{align}
where $\det(\mathcal{C}_2+\mathcal{I}_6)$ is the determinant of the
$6\times6$ second-order cumulant $\mathcal C_2 \eqdef \mathcal C_{2(\beta,\gamma)}$
plus the identity matrix $\mathcal{I}_6 \eqdef \delta_{\beta,\gamma}$.
In the context of the Hanbury-Brown--Twiss (HBT) effect
for the two-particle correlation function,
a similar formula for the size $R_s$ of the source function
is known~\cite{Blum:2017qnn,Bellini:2018epz,Blum:2019suo,Bellini:2020cbj}.
Although we do not explicitly perform diagrammatic calculations for the present study,
when the higher-order cumulants are sufficiently small,
the size of the higher-order corrections may be estimated as
\begin{align}
  H_A \eqdef \frac{N_A}{N_A^{(0)}} &= 1 +
    \int \biggl\{\prod_{i=1}^{A-1} \frac{
      d^6\bm\Aux_i \exp[-\frac12\bm{T}_i (\mathcal C_2 + \mathcal I_6) \bm{T}_i]}%
      {\sqrt{(2\pi)^6\det(\mathcal C_2 + \mathcal I_6)^{-1}}}
    \biggr\} \nonumber \\
  &\qquad\times
    \sum_{m=1}^\infty \frac{[\mathcal H(\{\bm{T}_i\}_{i=1}^{A-1})]^m}{m!} \nonumber\\
  &=1 + \mathcal O (\{\mathcal C_{\bm\alpha}\}_{|\bm\alpha|\ge3}),
  \label{eq:NA-higher-order-correction}
\end{align}
and thus
\begin{align}\label{eq:NA-factorized}
  N_A &= g_A N_p \biggl[\frac{8 N_p}{\sqrt{\det(\mathcal C_2 + \mathcal I_6)}}\biggr]^{A-1}\cdot
    [1 + \mathcal O (\{\mathcal C_{\bm\alpha}\}_{|\bm\alpha|\ge3})]
\end{align}
Here, it should be noticed that the magnitude of the higher-order correction $H_A$ is determined
by the higher-order cumulants $\{\mathcal C_{\bm\alpha}\}_{|\bm\alpha|\ge3}$
while the lower-order cumulants $\mathcal C_2$ do not directly contribute to the magnitude
because the lower orders only appear in the width of the normalized Gaussian kernel in Eq.~\eqref{eq:NA-higher-order-correction}.
In this way, in the expression of the light-nuclei yield $N_A=N_A^{(0)}H_A$,
we could successfully separate the contribution of the phase-space distribution
into its underlying Gaussian component $N_A^{(0)}$ (written in terms of the cumulants up to the second order)
and non-Gaussian corrections $H_A$ (essentially written in terms of the higher-order cumulants).

\subsection{Generalized yield ratio}

In Eq.~\eqref{eq:NA-factorized}, the lowest-order of the light-nuclei yields
of different mass numbers $A$ share a common structure of  $N_A \propto [\cdots]^{A-1}$,
where $[\cdots]$ is a common factor that does not depend on $A$
(if we assume a common light-nuclei size $\sigma_A \equiv \sigma$).
Motivated by this common structure,
we may consider particular combinations of the product of light-nuclei yields
to eliminate the background effect of the lowest order.
In general,
suppose the powers $\{p_k\}_{k=1}^{n_k}$ for the mass numbers $\{A_k\}_{k=1}^{n_k}$
satisfy the condition $\sum_{k=1}^{n_k} p_k (A_k-1) = 0$,
such a product can be constructed as
\begin{align}\label{eq:ratio-general}
  R_{A_1,\dots,A_{n_k}}^{p_1,\dots,p_{n_k}}
  &\eqdef N_p^{-\sum_{k=1}^{n_k} p_k} \prod_{k=1}^{n_k} N_{A_k}^{p_k} \nonumber \\
  &= \biggl(\prod_{k=1}^{n_k} g_{A_k}^{p_k}\biggr)
    [1 + \mathcal O (\{\mathcal C_{\bm\alpha}\}_{|\bm\alpha|\ge3})],
\end{align}
where the lowest order is genuinely determined by the statistical factors $g_{A_k}$.
A useful case would be $n_k=2$ with $p_A = -(B - 1)$ and $p_B = A - 1$
for mass numbers $A$ and $B$:
\begin{align}\label{ratio}
  R_{A,B}^{1-B,A-1} =
  \frac{N_p^{B-A} N_B^{A-1}}{N_A^{B-1}}
  &=\frac{g^{A-1}_B}{g_A^{B-1}}[1+\mathcal{O}(\{\mathcal{C}_{\bm{\alpha}}\}_{|\alpha| \ge 3})].
\end{align}
In fact, this corresponds to the yield ratio $R_{2,3}^{-2,1} = N_tN_p/N_d^2$~\cite{Sun:2017xrx} with $(A,B)$ being identified as (2,3)
and thus can be regarded as the generalization of the yield ratio
that roughly eliminates the background effect.
Therefore the generalized yield ratio~\eqref{eq:ratio-general}
gives the general formula for the light-nuclei yield ratios that are free
from the background effects of the underlying Gaussian component
of the background phase-space distribution.
It should be noted that
the effect of different numbers of protons and neutrons, $N_p\ne N_n$,
also cancels in the original yield ratio $N_tN_p/N_d^2$~\cite{Sun:2017xrx}
while it does not in e.g. $N_{^3\mathrm{He}} N_p/N_d^2$.
If we expect the same nature of the generalized ratio~\eqref{eq:ratio-general},
we may impose additional constraints on the nucleus species in the ratio
to balance the number of protons and neutrons.
More comprehensive studies on the isospin-asymmetry effect are left to the future study.

The dominant contribution from the second-order phase-space cumulants is canceled out in the ratios~\eqref{eq:ratio-general} and \eqref{ratio}.
Here, let us see what kind of information about the backgrounds has been physically eliminated.
The second-order phase-space cumulants are explicitly
written by the nucleon phase-space coordinates $(\bm{r},\bm{p})$ as
\begin{align}
  \mathcal{C}_2&=2\begin{pmatrix}
    \frac{\langle \bm{r}\bm{r}^\mathrm{T}\rangle}{\sigma^2} &
    \langle \bm{r}\bm{p}^\mathrm{T}\rangle \\
    \langle \bm{p}\bm{r}^\mathrm{T}\rangle &
    \sigma^2 \langle \bm{p}\bm{p}^\mathrm{T}\rangle
  \end{pmatrix},
\end{align}
where
\begin{align}
  \langle\bm{a}\bm{b}^\mathrm{T}\rangle &= \begin{pmatrix}
    \langle a_x b_x \rangle_c & \langle a_x b_y \rangle_c &\langle a_x b_z \rangle_c \\
    \langle a_y b_x \rangle_c & \langle a_y b_y \rangle_c &\langle a_y b_z \rangle_c \\
    \langle a_z b_x \rangle_c & \langle a_z b_y \rangle_c &\langle a_z b_z \rangle_c
  \end{pmatrix}. \nonumber
\end{align}
The diagonal components are the variances of coordinates $\langle r^2_i\rangle_c$ and momentum $\langle p^2_i \rangle_c$.
The former corresponds to the size of the fireball,
and the latter roughly corresponds to the effective temperature of the nucleon spectrum.
On the other hand, the non-diagonal components contain the correlation among different $r$'s and $p$'s.
These carry the information about the deformation of the fireball profiles in the coordinate and momentum space.
In particular, the coordinate--momentum correlation can be related to the radial flow,
which will be further discussed in Sec.~\ref{sec:example.radial}.
This means that all of these background effects are canceled in the yield ratios
so that the yield ratio becomes constant at the lowest order regardless of the background profiles.

Conversely, any non-constant behavior of the light-nuclei yield ratio
coming from the background contributions
can be attributed to the higher-order phase-space cumulants $\{\mathcal C_{\bm\alpha}\}_{|\bm{\alpha}|\ge3}$,
namely the non-Gaussian shape of the background phase-space distribution.
Such higher-order cumulants include $\langle r^3\rangle_c$ and
momentum $\langle p^4\rangle_c$ as well as the higher-order
space-momentum correlation $\langle r^2 p^2\rangle_c$.
Here, we conclude that the non-Gaussian components of the background phase-space distribution,
which are represented by the higher-order phase-space cumulants,
play an important role in understanding the non-constant behavior of the yield ratio observed in experiments.

\section{Examples of background phase-space distributions}\label{sec:example}

In Sec.~\ref{sec:formalism.yield}, we obtained the light-nuclei yield in terms of the phase-space cumulants
$\mathcal{C}_{\bm{\alpha}}$ with two assumptions:
(i) the same width parameter of the Wigner function $\sigma_A\equiv \sigma$
and (ii) the phase-space distribution being close to the Gaussian shape.
We then demonstrated that the yield ratio is a constant up to the second-order phase-space cumulants,
while the higher-order ones are important in the non-constant yield ratio.

To illustrate the result obtained in Sec.~\ref{sec:formalism}
and also to further investigate its physical properties,
we here calculate the light-nuclei yield ratio $N_t N_p /N_d^2$
with different background distributions and observe its behavior.
Note that the examples we employ in this section are toy models
for illustrative purposes and qualitative discussions,
which is far from the realistic one that can be used for the quantitative discussions.
For the comprehensive modeling of the heavy-ion collision reactions,
one should carefully tune the parameters of a realistic dynamical model
that are sensitive to the non-Gaussian part of the phase-space density
and compare the result with the non-monotonic behavior of the light-nuclei yield ratio in the experimental data.

\subsection{Gaussian phase-space distribution and size effects}\label{sec:example.gauss}
We first revisit the Gaussian example in Sec.~\ref{sec:coalescence.gauss},
where all the higher-order phase-space cumulants vanish
and reinterpret it using the general formula~\eqref{eq:NA4}.
The second-order phase-space cumulants of
the Gaussian phase-space distribution~\eqref{eq:dist-gauss} are
\begin{align}
  \mathcal C_2 =
  \begin{pmatrix}
    \frac{2R_s^2}{\sigma_A^2} \mathcal I_3 & 0 \\
    0 & 2\sigma_A^2 mT \mathcal I_3
  \end{pmatrix},
  \label{eq:example.gauss.C2}
\end{align}
where $\mathcal I_3$ is the $3\times3$ identity matrix.
The higher-order cumulants vanish: $\mathcal C_{\bm\alpha} = 0$ ($|\bm\alpha|\ge3$).
The yields \eqref{eq:gaussian-yields} can be now immediately obtained by using Eq.~\eqref{eq:NA4}.

While we have already observed that a particular yield ratio $N_tN_p/N_d^2$ becomes
a constant 4/9 under the assumption of a common light-nuclei size $\sigma$ in Sec.~\ref{sec:coalescence.gauss},
we now found that the generalized yield ratios~\eqref{eq:ratio-general} and \eqref{ratio}
also become constants for this Gaussian phase-space distribution.
Here, let us further estimate the effect of the size differences of $\sigma$
on the generalized yield ratio.
The light-nuclei ratio for the general mass number $A$ is written as
\begin{align}
  N_A &= g_A N_p^A \Bigl[\Bigl(R_s^2 + \frac{\sigma_A^2}2\Bigr)\Bigl(mT + \frac1{2\sigma_A^2}\Bigr)\Bigr]^{-3(A-1)/2}.
  \label{eq:esample.gauss.yield}
\end{align}
Now we represent the size-difference effect of $\sigma_A$
by quantities $\epsilon_A$ and $\eta_A$ defined as
\begin{align}
  R_s^2 + \frac{\sigma_A^2}2 &= \Bigl(R_s^2 + \frac{\sigma^2}2\Bigr)(1+\epsilon_A), \label{eq:example.gauss.epsA}\\
  mT + \frac1{2\sigma_A^2} &= \Bigl(mT + \frac1{2\sigma'^2}\Bigr)(1+\eta_A),
\end{align}
where $\sigma^2$ and $\sigma'^{-2}$ are typical sizes of $\sigma_A^2$ and $\sigma_A^{-2}$, respectively,
in the light-nuclei yield ratio.
For example, in the case of the yield ratio $N_tN_p/N_d^2$,
these can be specified as $\sigma^2 \eqdef (\sigma_d^2 + \sigma_t^2)/2 \simeq 3.8\ \text{fm}^2$
and $\sigma'^{-2} \eqdef (\sigma_d^{-2} + \sigma_t^{-2})/2 \simeq 0.30\ \text{fm}^{-2}$.
Given typical values $R_s^2 \sim 10^2\ \text{fm}^2$ and $mT \sim 1\cdot0.1\ \text{GeV}^2 \simeq 2.5\ \text{fm}^{-2}$,
the size-difference effects are estimated as
$\epsilon_{d,t} \sim \pm 6.3\times10^{-3}$ and $\eta_{d,t} \sim \mp 1.9\times10^{-2}$.
Therefore it is allowed to estimate the size-difference effect
in the generalized yield ratio~\eqref{eq:ratio-general} up to the leading order in $\epsilon_A$ and $\eta_A$:
\begin{align}
  &N_p^{-\sum_{k=1}^{n_k} p_k} \prod_{k=1}^{n_k} N_{A_k}^{p_k}
    =\biggl[ \prod_{k=1}^{} g_{A_k}^{p_k}\biggr] \nonumber \\
  &\times
    \biggl[1 - \frac32\sum_{k=1}^{n_k} p_k(A_k-1)(\epsilon_{A_k}+\eta_{A_k})
      + \mathcal O\bigl((\epsilon_A+\eta_A)^2\bigr) \biggr].
\end{align}
Using this formula,
the size-difference effect on, e.g., the yield ratio $N_tN_p/N_d^2$ can be estimated as
$N_tN_p/N_d^2 \sim (4/9)(1 - 7.5\times10^{-2})$,
which implies that the light-nuclei size difference is irrelevant
for the typical fireball size of $R_s \sim 10\ \text{fm}$.

\subsection{Woods--Saxon distribution and higher-order cumulants}\label{sec:example.ws}
\begin{figure}[htb]
  \includegraphics[width=0.48\textwidth]{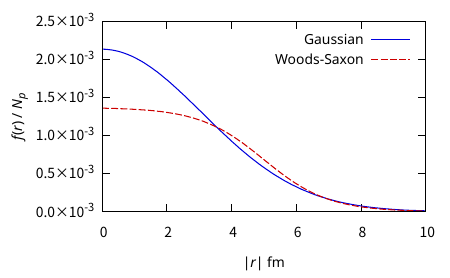}
  \caption{Comparison of Gaussian~\eqref{eq:dist-gauss} and the Woods--Saxon~\eqref{eq:example.ws.distribution} distributions
    of the same root-mean-square radius. The density is shown in units of $\text{fm}^{-3}$.
    The parameters of the Woods--Saxon distribution are chosen as $R_0=5.0\ \text{fm}$ and $a_0 = 1.0\ \text{fm}$.
    The width of Gaussian distribution is determined
    so that the root mean square of the radius matches that of the Woods--Saxon distribution:
    $R_s \eqdef \sqrt{\langle\bm{r}^2\rangle_\mathrm{WS}} = 3.099\ \text{fm}$.
    Assuming $N_p=1$, we obtain the normalization as $\rho_0 = 2.133\times10^{-3}\ \text{fm}^{-3}$
    and $\rho_\mathrm{WS} = 1.369\times10^{-3}\ \text{fm}^{-3}$ for Gaussian and the Woods--Saxon distributions, respectively.}
  \label{fig:example.ws.profile}
\end{figure}
To see how the higher-order phase-space cumulants affect the yield ratio
to produce non-constant behavior,
we here adopt a widely-used functional form,
the Woods--Saxon distribution.
For illustrative purposes, we here assume that the phase-space
distribution can be factorized into the momentum and position distributions,
as is done in Refs.~\cite{Sun:2018jhg,Sun:2020pjz}, where all the momentum
distributions are Gaussian.
The phase-space distribution is parametrized as
\begin{align}\label{eq:example.ws.distribution}
  f(\bm{r},\bm{p})&=\frac{\rho_{\mathrm{WS}}}{1+\exp\frac{r-R_0}{a_0}}
    \cdot \frac1{(2\pi mT)^{3/2}} \exp\Bigl(-\frac{\bm{p}^2}{2mT}\Bigr),
\end{align}
where the parameters $R_0$ and $a_0$ are the nuclear radius and the skin-thickness parameter
in the original context of the Woods--Saxon distribution for a nucleus.
It should be noted that these parameters do not have these particular meanings for the present purpose
but are rather treated as free parameters.
The parameter $\rho_{\mathrm{WS}}$ is determined by the normalization $N_p=\int d^3\bm{r}d^3\bm{p}f(\bm{r},\bm{p})$:
\begin{align}
  \rho_{\mathrm{WS}} &= \frac{N_p}{8\pi a_0^3 [-\PolyLog_3(-e^{R_0/a_0})]},
\end{align}
where
$\PolyLog_n(\lambda)\eqdef[1/(n-1)!]\int_0^\infty t^{n-1} dt/(e^t/\lambda-1)$ is the polylogarithm of the order $n$.
Fig.~\ref{fig:example.ws.profile} compares the radial profile of the Woods--Saxon distribution
with that of Gaussian distribution of the same root mean square of the fireball radius $\sqrt{\langle\bm{r}^2\rangle}$.

The characteristic function is
\begin{align}
  \phi(\bm{t})
  &= e^{-\frac12 mT\bm{t}_{\bm{p}}^2}
  \sum_{k=0}^\infty \frac{(\imag t_{\bm{r}})^{2k}}{(2k)!} \nonumber \\ &\quad
  \times \frac{a_0^{2k}}{2k+1} \frac{(2k+2)!\PolyLog_{2k+3}(-e^{R_0/a_0})}{2!\PolyLog_{3}(-e^{R_0/a_0})},
\end{align}
where $\bm{t}_{\bm{r}}$ and $\bm{t}_{\bm{p}}$ are the coordinate and momentum part of the vector $\bm{t}$,
and $t_{\bm{r}} \eqdef |\bm{t}_{\bm{r}}|$ is the norm.
The moments that contain an odd index vanish due to the symmetry.
The moments of even orders are
\begin{align}
  \langle r_x^{2l}r_y^{2m}r_z^{2n}\rangle
  &= \frac{a_0^{|\bm\alpha|}}{|\bm\alpha|+1} \frac{(|\bm\alpha|+2)!\PolyLog_{|\bm\alpha|+3}(-e^{R_0/a_0})}{2!\PolyLog_{3}(-e^{R_0/a_0})} \nonumber \\
  &\quad \times
    \Binomial{|\bm\alpha|/2}{l,m,n} \Binomial{|\bm\alpha|}{2l,2m,2n}^{-1},
\end{align}
where $|\bm\alpha|=2(l+m+n)$, and
$\tBinomial{n}{a,b,c} \eqdef n!/a!b!c!$ is the trinomial coefficient.
The higher-order cumulants of the Woods--Saxon distribution
that are constructed by these moments
have non-vanishing values.

\begin{figure}[htb]
  \includegraphics[width=0.48\textwidth]{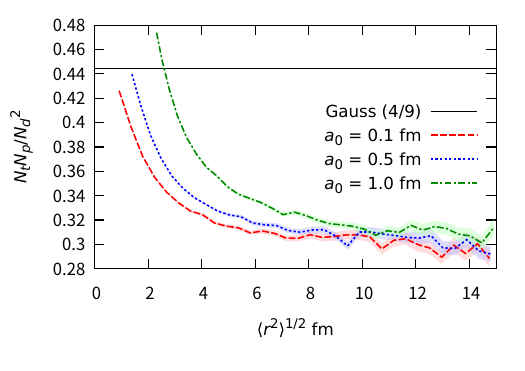}
  \caption{The light-nuclei yield ratio $N_t N_p/N_d^2$ as functions of
    the root-mean-square radius of the coordinate profile, $\langle r^2\rangle$.
    These are calculated for the common light-nuclei sizes $\sigma_d = \sigma_t = 1.59\ \text{fm}$.
    The freeze-out temperature and nucleon mass are
    assumed to be $T = 0.1\ \text{GeV}$ and $m = 0.939\ \text{GeV}$.
    The solid horizontal line represents the case with the Gaussian-shaped density profile.
    The other curves are the cases with the Woods--Saxon shaped density profile
    with different parameters of the skin thickness $a_0$.
    The band shows the statistical error of the Monte-Carlo integration.
    Oscillation is due to the statistical errors of the Monte-Carlo integration.}
  \label{figExample}
\end{figure}

\begin{figure}[htb]
  \includegraphics[width=0.48\textwidth]{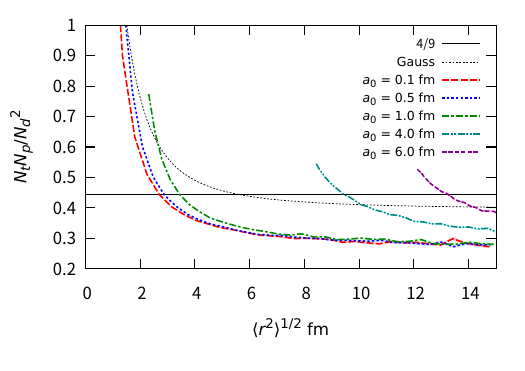}
  \caption{The light-nuclei yield ratio $N_t N_p/N_d^2$ for the physical light-nuclei sizes
    $\sigma_d = 2.26\ \text{fm}$ and $\sigma_t = 1.59\ \text{fm}$.
    The black dotted curve is the result of the Gaussian profile.
    The solid horizontal line shows the ideal value $g_t/g_d^2 = 4/9$.
    The other curves show the result for the Woods--Saxon shaped density profile.
    The setup is the same as in Fig.~\ref{figExample}.}
  \label{fig:physical-sigma}
\end{figure}

We first consider the case of a common light-nuclei size: $\sigma_d = \sigma_t = 1.59\ \text{fm}$
and obtain the light-nuclei yields using the Monte-Carlo integration.
In Fig.~\ref{figExample}, we plot the light-nuclei yield
ratio $N_t N_p/N_d^2$ with the Woods--Saxon distribution as a function of the variance $\langle \bm{r}^2\rangle_c = \sum_{i=1}^3\langle r^2_i\rangle_c$,
together with the one obtained from the Gaussian-shaped density profile~\eqref{eq:gaussian-yields} for comparison.
It shows that $N_tN_p/N_d^2$ is unchanged as a function of $\langle \bm{r}^2\rangle_c$ for the
Gaussian-shaped density distribution, whereas $N_tN_p/N_d^2$ decreases for
the Woods--Saxon one. This shows the importance of the non-Gaussianity of the
coordinate profile in the light-nuclei production.

We also check the behavior of the light-nuclei yield with the physical light-nuclei sizes.
In Fig.~\ref{fig:physical-sigma},
the light-nuclei yield ratio $N_tN_p/N_d^2$ is calculated with the physical values of
$\sigma_d = 2.26\ \text{fm}$ and $\sigma_t = 1.59\ \text{fm}$
for the Gaussian and the Woods--Saxon coordinate profiles.
With the physical light-nuclei sizes,
the Gaussian case also becomes non-constant
at the very small fireball sizes $R_s \sim \sigma_{d,t}$.
In fact, the size-difference effect~\eqref{eq:example.gauss.epsA}
becomes $\epsilon_{d,t} \simeq \pm5.91\times10^{-2}$ at $R_s = 3.0\ \text{fm}$,
and the ratio increases by 27\% as a consequence.
The qualitative behavior of the Woods--Saxon case does not change
compared to the case of the same sizes of deuterons and tritons.

\subsection{Double-Gaussian phase-space distribution}\label{sec:example.double}
\begin{figure}[htb]
  \centering
  \includegraphics[width=0.48\textwidth]{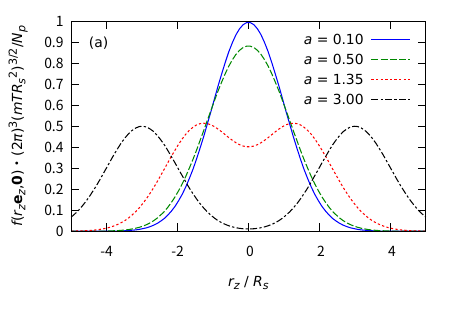}\\
  \includegraphics[width=0.48\textwidth]{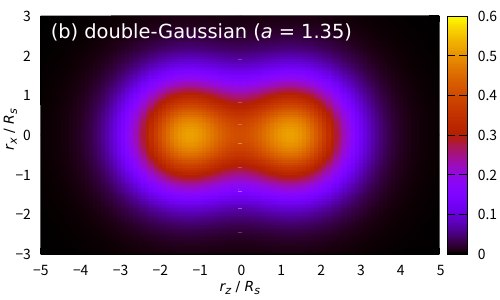}
  \caption{Panel (a) shows the profile of the double-Gaussian distribution for different $a = |\bm{r}_1-\bm{r}_2|/2R_s$.
    Panel (b) shows the $z$-$x$ profile of
    the phase-space distribution $f(\bm{r},0)(2\pi)^3(R_s^2mT)^{3/2}/N_p$
    for the case $a=1.35$ ($\alpha \sim 0.401$).}
  \label{fig:example.double.profile}
\end{figure}
In more realistic setups, we may consider structures inside the fireballs
caused by random nucleon distribution in the colliding nuclei.
The double-peak structure
by the incomplete baryon stopping depending on the collision energy
is also an interesting background.
We here consider the case of two Gaussian baryon-number hot spots at $\bm{r}_1$ and $\bm{r}_2$ inside the fireball.
We may choose $(\bm{r}_1+\bm{r}_2)/2$ as the coordinate origin
and $\bm{r}_1 - \bm{r}_2$ as the $r_z$-direction without loss of generality.
The phase-space distribution is written by
\begin{align}
  & f(\bm{r},\bm{p}) =N_p G(\bm{p};mT) \nonumber \\
  &\qquad \times \frac{
    G(\bm r - a R_s \bm{e}_z;R_s^2)
    + G(\bm r + a R_s \bm{e}_z;R_s^2)}2,
\end{align}
where $G(\bm{x};s^2) \eqdef (2\pi s^2)^{-3/2} \exp(-\bm{x}^2/2s^2)$
is Gaussian distribution of the variance $s^2$,
$\bm{e}_z = (0,0,1)^\mathrm{T}$ is the unit vector in the $r_z$-direction.
The parameters $R_s$ and $a = |\bm{r}_1 - \bm{r}_2|/2R_s$
specify the width and the relative separation of hot spots, respectively.
Panel (a) of Fig.~\ref{fig:example.double.profile}
shows the coordinate profile in the $r_z$-direction for different $a$'s,
and panel (b) shows an example profile for $a = 1.35$.

The cumulant-generating function becomes
\begin{align}
  \ln \phi(\bm{t})
    &= -\frac12 R_s^2 \bm{t}_{\bm{r}}^2 - \frac12 mT \bm{t}_{\bm{p}}^2 + \ln \cos (aR_s t_z), \nonumber \\
    &= -\frac12 R_s^2 \bm{t}_{\bm{r}}^2 - \frac12 mT \bm{t}_{\bm{p}}^2 \nonumber \\ &\qquad
      + \sum_{k=1}^{\infty} \frac{(it_z)^{2k}}{(2k)!} 2^{2k} (2^{2k} - 1)\frac{B_{2k}}{2k} (aR_s)^{2k},
\end{align}
where $\bm{t}_{\bm{r}}$, $\bm{t}_{\bm{p}}$, and $\bm{t}_z$ are
the conjugate variables of $\bm{r}$, $\bm{p}$, and $r_z$, respectively,
and $B_{2k}$ are the Bernoulli numbers: $B_2 = 1/6$, $B_4 = -1/30$, $B_6 = 1/42$, $\ldots$.
The cumulants are the same as Sec.~\ref{sec:example.gauss} except for $\langle r_z^{2k}\rangle_c$:
\begin{align}
  \langle r_z^2\rangle_c &= R_s^2 (1 + a^2), \\
  \langle r_z^{2k}\rangle_c / R_s^{2k} &= 2^{2k} (2^{2k}-1) \frac{B_{2k}}{2k} a^{2k}, & (k &\ge 2).
\end{align}

\begin{figure}[htb]
  \centering
  \includegraphics[width=0.48\textwidth]{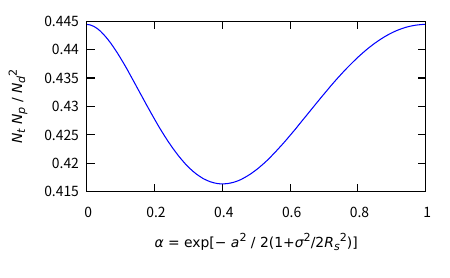}
  \caption{Yield ratio as a function of $\alpha$.
    The value $\alpha=1$ corresponds to $a=0$, and $\alpha=0$ corresponds to $a=\infty$.}
  \label{fig:example.double.ratio}
\end{figure}
The analytical result of the light-nuclei yield is
\begin{align}
  N_A &= g_A N_p^A \Bigl[\Bigl(R_s^2 + \frac{\sigma_A^2}2\Bigr)\Bigl(mT + \frac1{2\sigma_A^2}\Bigr)\Bigr]^{-3(A-1)/2} \nonumber \\
  &\times \frac1{2^A} \sum_{r=0}^A \Binomial{A}{r} \alpha^{A[1-(2r/A-1)^2]},
  \label{eq:example.double.yield}
\end{align}
where $\tBinomial Ar \eqdef A!/r!(A-r)!$ is the binomial coefficient,
and $\alpha \eqdef \exp\{-a^2/[2(1+ \sigma_A^2/2R_s^2)]\}$
[see Appendix~\ref{app:example.double.yield} for the derivation of Eq.~\eqref{eq:example.double.yield}].
With the assumption of a common nuclei size $\sigma_A\equiv\sigma$, the light-nuclei yield ratio is
\begin{align}
  \frac{N_tN_p}{N_d^2} &= \frac49 \frac{1+3\alpha^{8/3}}{(1+\alpha^2)^2}.
\end{align}
Fig.~\ref{fig:example.double.ratio} shows the yield ratio as a function of $\alpha \in [0,1]$.
It becomes the Gaussian value $4/9$ for $\alpha=0$ and $1$,
which correspond to the case $a=\infty$ and $0$, respectively.
In the case of $\alpha=1$ ($a=0$), this is simply because the phase-space distribution is reduced to a single Gaussian distribution.
An interesting case is $\alpha=0$ ($a=\infty$) where we naturally expect the Gaussian value $4/9$
because the phase-space distribution becomes the sum of two separate Gaussian distributions.
Nevertheless, the higher-order cumulants do not vanish for these two separate Gaussian distributions.
This means that the non-vanishing higher-order cumulants do not always lead to the deviation of the yield ratio.
We see that the yield ratio becomes smaller with the intermediate values of $\alpha$
where there is spatial structures in a fireball.
The yield ratio takes the minimum value at $\alpha = (\sqrt[3]{3\sqrt{33}+17}-\sqrt[3]{3\sqrt{33}-17}-1)^{3/2}/\sqrt{27}\simeq 0.401$,
which corresponds to $a \simeq 1.35 \sqrt{1+\frac{\sigma^2}{2R_s^2}}\sim 1.35$ ($\sigma^2 \ll R_s^2$).
The case for $a=1.35$ is shown by the red dotted line in panel (a) of Fig.~\ref{fig:example.double.profile}.
The profile in the $z$-$x$ plane shown in panel (b) of Fig.~\ref{fig:example.double.profile} actually corresponds to this case $a=1.35$.

\subsection{Radial flow and coordinate--momentum correlation}\label{sec:example.radial}
So far, we have been assuming the factorization of the coordinate and momentum profiles for illustrative purposes.
Nevertheless, our formula for the light-nuclei yield~\eqref{eq:NA4} is also capable of dealing
with the non-factorizable form of the phase-space distribution.
In the realistic descriptions of the heavy-ion collision reactions,
the coordinate--momentum correlation becomes important for the light-nuclei production~\cite{Zhao:2020irc,Zhao:2018lyf}.
Here, we combine the Gaussian phase-space distribution~\eqref{eq:dist-gauss}
with the blast-wave type of flow configuration in the transverse plane:
\begin{align}
  \bm{v}(\bm{r}) &= (v_x, v_y, v_z)^\mathrm{T} \eqdef \frac1{R_s}(r_x, r_y, 0)^\mathrm{T}.
  \label{eq:example.radial.velocity}
\end{align}
The phase-space distribution under the flow velocity $\bm{v}(\bm{r})$ in the non-relativistic regime becomes
\begin{align}
  f(\bm{r},\bm{p}) &= \frac{\rho_0}{(2\pi mT)^{3/2}} e^{-\frac{\bm{r}^2}{2R_s^2}}
    \exp\Bigl(- \frac{m}{2T}\Bigl[\frac{\bm{p}}m-\bm{v}(\bm{r})\Bigr]^2\Bigr),
  \label{eq:example.radial.distribution}
\end{align}
where $\rho_0 = N_p/(2\pi R_s^2)^{3/2}$ is the same as in Sec.~\ref{sec:coalescence.gauss}.
Because the exponent is a quadratic form of $\bm{r}$ and $\bm{p}$,
this phase-space distribution is still Gaussian.
The second-order phase-space cumulants are obtained as
\begin{align}
  \mathcal C_2 &= 2\begin{pmatrix}
    \frac{R_s^2}{\sigma_A^2} \mathcal I_3
    & \quad R_s m\mathcal I_\perp \\
    R_s m\mathcal I_\perp
    & \quad \sigma_A^2 (mT \mathcal I_3 + m^2 \mathcal I_\perp)
  \end{pmatrix},
\end{align}
where $\mathcal I_\perp\equiv \diag(1,1,0)$ is the projection to the transverse plane.
The coordinate--momentum correlations
$\langle r_x p_x\rangle_c = \langle r_y p_y\rangle_c = R_s m$
or $\langle \bm{r}_\perp\cdot\bm{p}_\perp\rangle_c \equiv\langle r_xp_x + r_yp_y\rangle_c = 2R_s m$
is associated with the strength of the radial flow.
The higher-order phase-space cumulants $\{\mathcal C_{\bm\alpha}\}_{|\bm\alpha|\ge3}$ vanish.
The light-nuclei yield is
\begin{align}
  N_A
  &= g_A N_p^A \Bigl[\Bigl(R_s^2 + \frac{\sigma_A^2}2\Bigr)\Bigl(mT + \frac1{2\sigma_A^2}\Bigr)\Bigr]^{-(A-1)/2} \nonumber \\
  &\quad\times \Bigl[\Bigl(R_s^2 + \frac{\sigma_A^2}2\Bigr)\Bigl(mT + \frac1{2\sigma_A^2}\Bigr) + \frac{m^2 \sigma_A^2}2 \Bigr]^{-(A-1)}.
\end{align}
Finally, the light-nuclei yield ratio becomes constant $N_tN_p/N_d^2 = 4/9$
as a consequence of the vanishing higher-order phase-space cumulants.
This means that even the effect of the background radial flow of the blast-wave type
does not affect the light-nuclei yield ratio with the Gaussian coordinate profile.

\subsection{Elliptic flow}\label{sec:example.elliptic}
\begin{figure}[htb]
  \centering
  \includegraphics[width=0.4\textwidth]{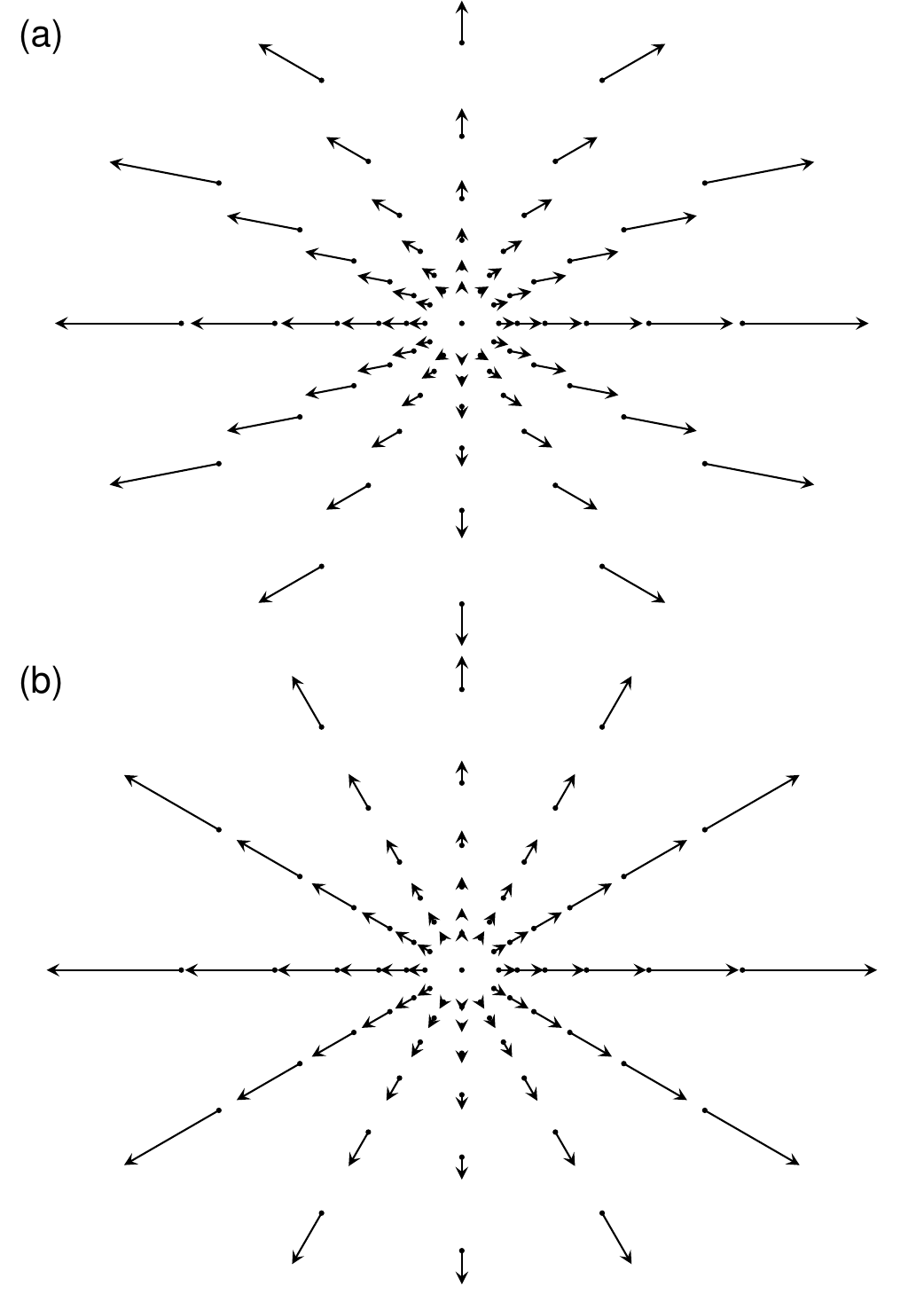}
  \caption{Panels (a) and (b) show the flow configurations
  of Eqs.~\eqref{eq:flow-elliptic} and \eqref{eq:flow-blast2}, respectively.
  The anisotropy parameters are $\varepsilon = 0.5$ and $u_2 = 0.3$, respectively.}
  \label{fig:example.elliptic.flow}
\end{figure}
In Sec.~\ref{sec:example.radial}, we have introduced isotropic radial flow.
With the current framework, it is even possible to introduce anisotropic flow.
The way to introduce the anisotropic flow in the phase-space distribution is not unique.
In this subsection, we first consider the following simple flow configuration
for the phase-space distribution~\eqref{eq:example.radial.distribution}:
\begin{align}\label{eq:flow-elliptic}
  \bm{v}(\bm{r}) &= \frac1{R_s}\bigl(r_x(1+\varepsilon), r_y(1-\varepsilon), 0\bigr)^\mathrm{T},
\end{align}
where $\varepsilon$ is a parameter to control the anisotropy.
The flow profile is shown in panel (a) of Fig.~\ref{fig:example.elliptic.flow}.
The corresponding phase-space distribution is still Gaussian
as is in the case of the isotropic radial expansion~\eqref{eq:example.radial.velocity}.
The second-order cumulants become
\begin{align}
  \mathcal C_2 &= 2\begin{pmatrix}
    \frac{R_s^2}{\sigma_A^2} \mathcal I_3
    & R_s m (\mathcal I_\perp + \varepsilon\lambda_3) \\
    R_s m(\mathcal I_\perp + \varepsilon\lambda_3)
    & \sigma_A^2 [mT \mathcal I_3 + m^2 (\mathcal I_\perp + \varepsilon\lambda_3)^2]
  \end{pmatrix},
\end{align}
where $\lambda_3 = \diag(1, -1, 0)$,
and the higher-order phase-space cumulants vanish.

Using the second-order cumulants,
one can relate the asymmetry of the coordinate--momentum correlations
with the anisotropy parameter:
\begin{align}
  \frac{\langle r_xp_x-r_yp_y\rangle_c}{\langle r_xp_x + r_yp_y\rangle_c}
  &= \frac{\langle\bm{r}^\mathrm{T} \lambda_3 \bm{p}\rangle_c}{\langle\bm{r}^\mathrm{T} \mathcal I_\perp \bm{p}\rangle_c}
  = \varepsilon.
\end{align}
The momentum eccentricity is expressed by the momentum cumulants:
\begin{align}
  \varepsilon_p
  &= \frac{\langle p_x^2 - p_y^2\rangle}{\langle p_x^2 + p_y^2\rangle}
  = \frac{\langle\bm{p}^\mathrm{T} \lambda_3 \bm{p}\rangle_c}{\langle\bm{p}^\mathrm{T} \mathcal I_\perp \bm{p}\rangle_c}
  = \frac{\varepsilon}{1+\varepsilon^2 + \frac Tm}.
\end{align}
Although the elliptic-flow coefficient $v_2$ cannot be
directly calculated by the second-order phase-space cumulants,
it can be analytically calculated (see Appendix~\ref{app:example.elliptic.flow}):
\begin{align}
  v_2 &\eqdef \langle\cos2\phi_p\rangle
  = \frac{\sqrt{\rho_+}-\sqrt{\rho_-}}{\sqrt{\rho_+}+\sqrt{\rho_-}},
  \label{eq:example.elliptic.flow}
\end{align}
where $\phi_p \eqdef \arg(p_x + \imag p_y)$, and
$\rho_\pm \eqdef 1 + \frac mT (1\pm\varepsilon)^2$.

\begin{figure}[htb]
  \centering
  \includegraphics[width=0.44\textwidth]{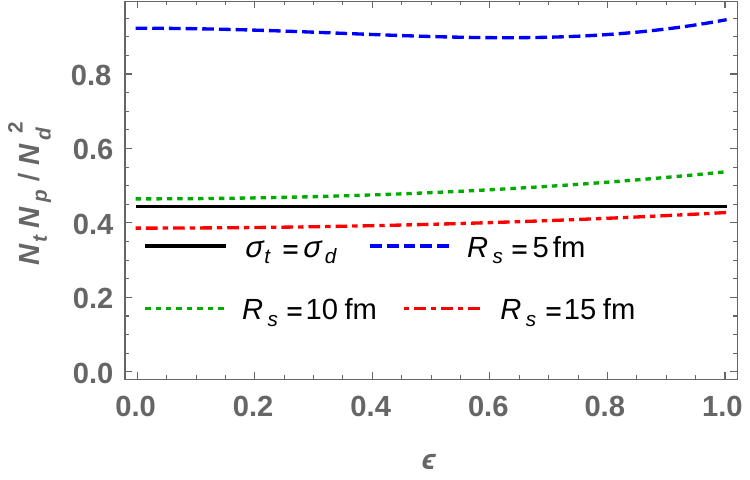}
  \caption{Light-nuclei yield ratio for the elliptic flow configuration~\eqref{eq:flow-elliptic}
    with the physical light-nuclei sizes $\sigma_t$ and $\sigma_d$.
    The black solid line shows 4/9 for the same size case $\sigma_t = \sigma_d$.
    The other curves show the results for different fireball sizes, $R_s$.}
  \label{fig:example.elliptic.epsilon}
\end{figure}
The light-nuclei yield is
\begin{align}
  N_A
  &= g_A N_p^A \Bigl[\Bigl(R_s^2 + \frac{\sigma_A^2}2\Bigr)\Bigl(mT + \frac1{2\sigma_A^2}\Bigr)\Bigr]^{-\frac{A-1}2} \nonumber \\
  &\times \Bigl[\Bigl(R_s^2 + \frac{\sigma_A^2}2\Bigr)\Bigl(mT + \frac1{2\sigma_A^2}\Bigr) + \frac{m^2 \sigma_A^2}2 (1+\varepsilon)^2 \Bigr]^{-\frac{A-1}2} \nonumber \\.
  &\times \Bigl[\Bigl(R_s^2 + \frac{\sigma_A^2}2\Bigr)\Bigl(mT + \frac1{2\sigma_A^2}\Bigr) + \frac{m^2 \sigma_A^2}2 (1-\varepsilon)^2 \Bigr]^{-\frac{A-1}2}.
\end{align}
As expected from the fact that the phase-space distribution is Gaussian,
the light-nuclei yield ratio becomes the constant $N_tN_p/N_d^2 = 4/9$ for the same light-nuclei sizes $\sigma_t = \sigma_d$.
This means that the background effect of the elliptic flow of the form~\eqref{eq:flow-elliptic}
is also canceled out in the yield ratio.
The size-difference effect is checked with the physical light-nuclei sizes in Fig.~\ref{fig:example.elliptic.epsilon}.
Although we observe the deviation of the yield ratio from the ideal value 4/9,
the deviation does not depend on the anisotropy parameter $\varepsilon$,
which means that the value of the yield ratio is mostly determined
by that of the vanishing-flow case ($\varepsilon = 0$).

\subsection{Blast-wave--type anisotropic flow}\label{sec:example.blastwave}

In the previous subsection, we have studied a naive elliptic-flow configuration~\eqref{eq:flow-elliptic}.
We may consider another type of the flow configuration for the phase-space distribution~\eqref{eq:example.radial.distribution},
the blast-wave--type anisotropic flow~\cite{Huovinen:2001cy}:
\begin{align}\label{eq:flow-blast2}
  \bm{v}(\bm{r}) &= \frac1{R_s}(r_x, r_y, 0)^\mathrm{T} (1+ 2u_2 \cos2\phi_s),
\end{align}
where $\phi_s \eqdef \arg(r_x + \imag r_y)$ is the spatial azimuthal angle,
and $u_2$ is a parameter to control the anisotropy.
The flow configuration is illustrated in panel (b) of Fig.~\ref{fig:example.elliptic.flow}.
Even though this second type of the flow configuration is often used in the context of the blast-wave model,
it should be noted that this is not necessarily be the flow configuration generated in the realistic setups.
As the elliptic flow is mainly created by the pressure gradient of the initial almond shape,
it is not necessarily in the radial direction from the fireball center.
The flow configuration of the previous subsection~\eqref{eq:flow-elliptic} shown in panel (a) of Fig.~\ref{fig:example.elliptic.flow}
may better capture the qualitative behavior naively.
Nevertheless, it is instructive to compare the behaviors of these two different phenomenological flow configurations
to check the robustness of the elliptic-flow effect on the yield ratio.

\begin{figure}[htb]
  \centering
  \includegraphics[width=0.48\textwidth]{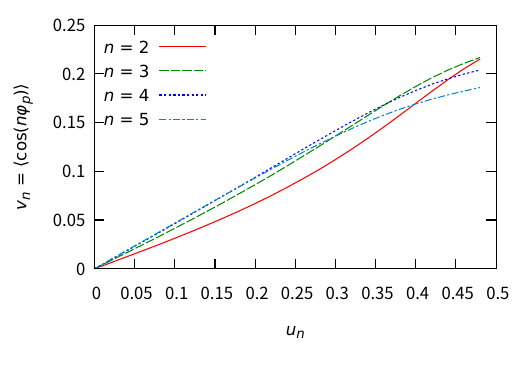}
  \caption{Azimuthal anisotropy $v_n = \langle\cos n\phi\rangle$ of the order $n$
    as a function of the anisotropy parameter $u_n$ in the phase-space distribution
    of the blast-wave--type flow configuration Eqs.~\eqref{eq:flow-blast2} and \eqref{eq:flow-blast3}.}
  \label{fig:example.elliptic.vn}
\end{figure}
\begin{figure}[htb]
  \centering
  \includegraphics[width=0.48\textwidth]{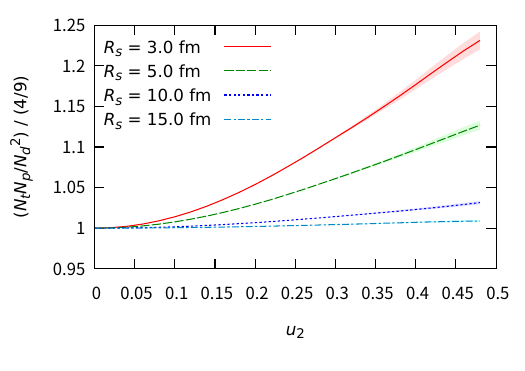}
  \caption{Light-nuclei ratios $N_tN_p/N_d^2$
    as functions of the elliptic anisotropy parameter $u_2$ \eqref{eq:flow-blast2} in the phase-space distribution.
    We here assumed the fixed $\sigma = 2.0\ \text{fm}$.
    The light-nuclei yields \eqref{eq:coale-1} are calculated
    by the Monte-Carlo integration with importance sampling (see Appendix~\ref{app:importance-sampling}).
    The values are normalized by the ideal value $4/9$.
    Different lines show the results for different fireball sizes, $R_s$.
    The band shows the statistical errors of the Monte-Carlo integration.
    The freeze-out temperature and nucleon mass are
    set as $T = 0.1\ \text{GeV}$ and $m = 0.939\ \text{GeV}$.}
  \label{fig:example.elliptic.ratio-size}
\end{figure}
It is difficult to perform analytical calculations for this flow configuration
due to $\cos2\phi_s = \cos^2\phi_s - \sin^2\phi_s = (r_x^2 - r_y^2)/(r_x^2 + r_y^2)$ in the exponent
of the distribution function.
Instead, the results are obtained numerically.
The red solid line in Fig.~\ref{fig:example.elliptic.vn}
shows the numerical result for the elliptic flow $v_2$ as a function of the parameter $u_2$.
Fig.~\ref{fig:example.elliptic.ratio-size} shows the light-nuclei yield ratios
for different fireball sizes, $R_s$, as functions of the anisotropy parameter $u_2$.
We here assumed the fixed light-nuclei size, $\sigma = 2.0\ \text{fm}$.
There is a deviation of the yield ratio from the ideal value $4/9$
unlike the case for the flow configuration of the previous subsection~\eqref{eq:flow-blast2}.
This means that the light-nuclei yield ratio
is not just determined by the values of the azimuthal anisotropy
but largely depends on the detailed flow configuration of the evolving systems.
We also find that the deviation of the yield ratio from $4/9$
is more significant for smaller sizes of fireballs,
which are closer to the light-nuclei size $\sigma$.

\begin{figure}[htb]
  \centering
  \includegraphics[width=0.48\textwidth]{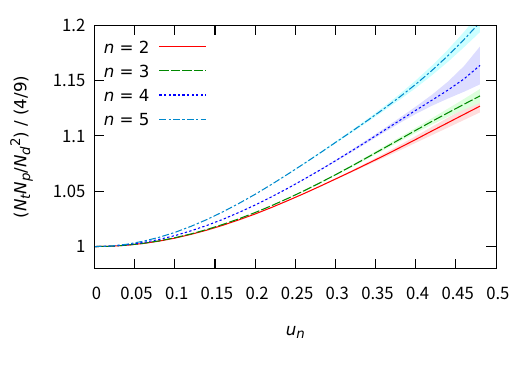}
  \caption{Light-nuclei ratios $N_tN_p/N_d^2$
    as functions of different orders of the anisotropy parameter $u_n$
    in the blast-wave--type phase-space distribution.
    The fireball and the light-nuclei sizes are fixed as $R_s = 5.0\ \text{fm}$ and $\sigma = 2.0\ \text{fm}$.
    Different lines show the results for different orders, $n$, of the anisotropy.
    The band shows the statistical errors of the Monte-Carlo integration (see Appendix~\ref{app:importance-sampling}).
    The other setup is the same as in Fig.~\ref{fig:example.elliptic.ratio-size}.}
  \label{fig:example.elliptic.ratio-order}
\end{figure}
We may also consider the general order of the anisotropic flow:
\begin{align}\label{eq:flow-blast3}
  \bm{v}(\bm{r}) &= \frac1{R_s}(r_x, r_y, 0)^\mathrm{T} (1+ 2u_n \cos n\phi_s),
\end{align}
where $u_n$ is the $n$-th order anisotropy parameter.
In Fig.~\ref{fig:example.elliptic.vn}, we confirm that the phase-space distribution
induces the azimuthal anisotropy $v_n = \langle\cos n\phi\rangle$.
The resulting light-nuclei yield ratio is shown in Fig.~\ref{fig:example.elliptic.ratio-order}.
We observe that the higher-order anisotropy creates more deviation of the light-nuclei yield from the ideal value 4/9.
This is because the higher-order anisotropy creates smaller structures in the phase-space distribution of the flow profile.

\section{Conclusion and Outlook}\label{conclusion}

Within the framework of the coalescence model, we have studied the effect of non-critical fluctuation
on the yield ratio of the light-nuclei.
To achieve the conceptual clarity of the dependence on the phase-space distribution $f(\bm{r},\bm{p})$,
we employed the characteristic function of the distribution,
with which we derived the expression of the light-nuclei yield
in terms of the cumulants of the phase-space distribution in Sec.~\ref{sec:formalism}.
Through the characteristic function of the phase-space distribution,
we decomposed the phase-space distribution into different orders of the \textit{phase-space cumulants}.
The second-order phase-space cumulants correspond to the underlying Gaussian component of the phase-space distribution,
whereas the higher-order cumulants correspond to the non-Gaussian distortion of the phase-space distribution.
This enabled us to study the contribution of the Gaussian part and the various non-Gaussian parts separately in a systematic way.
We found that the leading terms (up to the second-order of phase-space cumulants) in the light-nuclei yield
with arbitrary mass numbers $A$ share the same structure.
Thus, under the approximation of the same light-nuclei radii $\sigma_A \equiv \sigma$,
we constructed the generalized light-nuclei yield ratio
\begin{align*}
  R_{A,B}^{1-B,A-1} &= \frac{N_B^{A-1} N_p^{B-A}}{N_A^{B-1}}
\end{align*}
so that the leading-order terms are canceled out.
The above yield ratio can be regarded as the generalization
of the yield ratio $R_{2,3}^{-2,1}=N_tN_p/N_d^2$~\cite{Sun:2017xrx}.

The canceled leading-order terms contain the second-order phase-space cumulants $\mathcal C_2$
that represent the underlying Gaussian shape in the phase space.
Specifically, $\mathcal{C}_2$ consists of the variance in the
coordinate space $\langle r_i r_j \rangle_c$ and the momentum space
$\langle p_i p_j \rangle_c$ as well as the coordinate--momentum
correlation $\langle r_i p_j \rangle_c$, which are respectively related to
the fireball size and shape,
the broadening of the momentum at the freeze-out surface (i.e., roughly the kinetic freeze-out temperature),
and the expansion rate $\theta\eqdef\partial_\mu u^\mu$ induced by the radial flows and the longitudinal expansion, respectively.
Here, it is remarkable that the second-order cumulants can be canceled out
in the light-nuclei ratios.
In particular, we can see the global expansion of the fireball,
whose dominant component can be expressed
by the mixed second-order cumulants $\langle\bm{r}\cdot\bm{p}\rangle_c$,
is canceled in the light-nuclei ratio.
This means that these important effects of the backgrounds
are already roughly canceled in the light-nuclei yield ratio,
which supports the fact that the light-nuclei yield ratio is a desirable observable
to search the critical point.

In understanding any deviation and the variation of
the light-nuclei yield ratio from the ideal values $g^{A-1}/g^{B-1}_A$,
the higher-order cumulants, i.e., the non-Gaussianity, of the phase-space distribution plays a significant role.
The higher-order phase-space cumulants $\{\mathcal{C}_{\bm\alpha}\}_{|\bm{\alpha}|\ge3}$,
such as the skewness, kurtosis of the coordinate and momentum as well
as the coskewness and cokurtosis between them, become the dominant factors of the background effect in the ratio.
This non-Gaussianity of the phase-space shape may arise from various physics,
including the initial Woods--Saxon profile, the
non-trivial time evolution of the speed of sound $c_s(t)$, the
long-lived resonance decay that contributes to the long exponential tails of
the phase-space, and also the event-by-event modification from the
critical fluctuations.

To reveal the qualitative feature of the background contributions to the light-nuclei yield ratio in more detail,
we investigated several parametrized background phase-space distributions
based on the perspective obtained by the characteristic function
and the phase-space cumulants.
In Sec.~\ref{sec:example.gauss}, we first considered the Gaussian spatial profile
and obtained the light-nuclei yield \eqref{eq:esample.gauss.yield} for the general mass number $A$
with the spherical harmonic-oscillator Wigner function.
We also estimated the effect of different nuclei sizes $\sigma_t\ne\sigma_d$.
In Sec.~\ref{sec:example.ws},
we observed the effect of the higher-order phase-space cumulants
using the Woods--Saxon spatial profile of the phase-space distribution.
We also checked the effect of the size difference between tritons and deuterons.
In Sec.~\ref{sec:example.double},
we obtained the analytic expression of the light-nuclei yield ratio
for another non-Gaussian phase-space distribution with two hot spots,
where we found that the yield ratio takes a minimum
at an intermediate distance between the two hot spots.
In Sec.~\ref{sec:example.radial},
we argued that the radial flow in the Gaussian profile
introduced by a naive non-relativistic blast-wave--type flow configuration
does not affect the yield ratio.
In this sense, the yield ratio is also unaffected against the coordinate--momentum correlations
in the phase-space distribution caused by some types of radial expansion.
In Secs.~\ref{sec:example.elliptic} and \ref{sec:example.blastwave},
we introduced the elliptic flows with two different flow configurations.
We showed that a naive elliptic flow configuration~\eqref{eq:flow-elliptic}
is still a Gaussian phase-space distribution so results in the constant yield ratio of 4/9.
On the other hand, we found that another blast-wave--type configuration~\eqref{eq:flow-blast2}
may largely affect the light-nuclei yield ratio.
The size of the effect depends on the fireball size;
the effect is mostly negligible when $R_s = 15\ \text{fm}$
while it is significant in the smaller systems.
We also checked different orders of the blast-wave--type anisotropic flows
to find that the higher orders have larger effects on the light-nuclei yield ratio.

For future studies, the quantitative analysis with more realistic setups would be important.
In this study, a part of the effect of the radial expansion and flows is turned out to be canceled in the yield ratio,
which implies that the variation of the yield ratio carries the information about more details of the dynamics.
This suggests that the future quantitative analysis of various related physics
would provide promising insights into the experimental measurements,
which requires comprehensive studies.
The related physics include
the complex evolution of fluids (possibly related to the non-trivial initial conditions
and the subsequent hydrodynamic evolution with realistic EoS),
the contribution of the non-Gaussian profile from the long-lived resonance,
and, most importantly, the critical fluctuations.
In the realistic setup of the high-energy heavy-ion collisions,
the event-by-event fluctuations are also important ingredients,
which can also affect the light-nuclei yields.
The presented analysis by the characteristic function and the phase-space cumulants
can also be naturally extended to include the effect of the critical fluctuations
and other effects not considered in the present analysis such as the deviation of the Wigner function from the Gaussian
and the isospin asymmetry of the phase-space distribution.
Other challenges would include the extension of the analysis for the boost-invariant backgrounds
associated with dynamically expanding coordinates of $\tau = \sqrt{t^2-z^2}$ and $\eta_s = \tanh^{-1}(z/t)$, i.e., the Milne coordinates,
and the relativistic form of the phase-space distribution,
as well as the effect of the finite rapidity acceptance in experiments.

In this paper, for the critical-point search,
we have focused on the yield ratios where the non-critical background effect
becomes minimal but is still non-negligible.
Nevertheless, our analysis may also be naturally extended
for the observables of the light-nuclei production
that largely affected by the background effects,
such as the coalescence factor $B_A(p_\mathrm{T})$,
which has attracted attention recently~\cite{Wang:2020zaw}.

\section*{Acknowledgments}
We would like to thank the fruitful discussion with Kai-Jia Sun,
Xiaofeng Luo, Dingwei Zhang,  Christopher Plumberg and Shujun Zhao.
This work is supported by the NSFC under grant No. 12075007 and No. 11947236 as well as the China Postdoctoral Science Foundation under Grant No. 2020M680184.
We also gratefully acknowledge the extensive computing resources provided
by the Super-computing Center of Chinese Academy of Science (SCCAS),
Tianhe-1A from the National Supercomputing Center in Tianjin, China
and the High- performance Computing Platform of Peking University.

\appendix

\section{Light-nuclei yields with a simple Gaussian phase-space distribution}\label{app:simple-gauss}
The light-nuclei yield for the mass number $A$ under the phase-space distribution~\eqref{eq:dist-gauss} is written as
\begin{align}
  N_A &= g_A 8^{A-1} \left[\frac{N_p}{(2\pi R_s^2)^{3/2}(2\pi m T)^{3/2}}\right]^A
    \int\biggl[\prod_{i=1}^A d^3\bm{r}_i d^3\bm{p}_i\biggr] \nonumber \\
    &\quad\times e^{-\frac{1}{2R_s^2}\sum_{i=1}^A \bm{r}_i^2 - \frac1{2mT} \sum_{i=1}^A \bm{p}_i^2} \nonumber \\
    &\quad\times e^{-\frac1{\sigma_A^2} \sum_{i=1}^{A-1} \bm{R}_i^2 - \sigma_A^2 \sum_{i=1}^{A-1} \bm{P}_i^2}.
\end{align}
We can actually easily calculate this integration using the transformation introduced in Sec.~\ref{sec:phasevariable}.
We rewrite $(\bm{r}_i,\bm{p}_i)$ with $(\bm{R}_i,\bm{P}_i)$ using the relation~\eqref{eq:zZ2}:
\begin{align}
  N_A &= g_A 8^{A-1} \left[\frac{N_p}{(2\pi R_s^2)^{3/2}(2\pi m T)^{3/2}}\right]^A
    \int\biggl[\prod_{i=1}^A d^3\bm{R}_i d^3\bm{P}_i\biggr] \nonumber \\
    &\quad\times e^{-\frac1{2R_s^2} \bm{R}_A^2 - \frac1{2mT} \bm{P}_A^2} \nonumber \\
    &\quad\times e^{-(\frac{1}{2R_s^2}+\frac1{\sigma_A^2})\sum_{i=1}^{A-1} \bm{R}_i^2 - (\frac1{2mT} + \sigma_A^2) \sum_{i=1}^{A-1} \bm{P}_i^2}.
\end{align}
Now, we are ready to perform the Gaussian integrations to obtain the result:
\begin{align}
  N_A &= g_A 8^{A-1} \left[\frac{N_p}{(2\pi R_s^2)^{3/2}(2\pi m T)^{3/2}}\right]^A \nonumber \\
    &\quad\times(2\pi R_s^2)^{3/2} (2\pi mT)^{3/2} \nonumber \\
    &\quad\times\Bigl[
      2\pi \Bigl(\frac1{R_s^2} + \frac2{\sigma_A^2}\Bigr)^{-1} \cdot
      2\pi \Bigl(\frac1{mT} + 2\sigma_A^2\Bigr)^{-1}\Bigr]^{\frac{3(A-1)}2} \nonumber \\
  &= g_A N_p^A \Bigl[
      \Bigl(R_s^2 + \frac{\sigma_A^2}2\Bigr)\cdot
      \Bigl(mT + \frac1{2\sigma_A^2}\Bigr)\Bigr]^{-\frac{3(A-1)}2}.
\end{align}

\section{Derivation of Eq.~\eqref{eq:example.double.yield}: The light-nuclei yield for the double-Gaussian distribution}%
\label{app:example.double.yield}

Here, the derivation of Eq.~\eqref{eq:example.double.yield} is explained.
First, we define the distribution functions for each hot spot:
\begin{align}
  f(\bm{r},\bm{p}) &= \frac12[f_+(\bm{r},\bm{p}) + f_-(\bm{r},\bm{p})], \\
  f_\pm(\bm{r},\bm{p})
    &\eqdef N_p G(\bm r \mp a R_s \bm{e}_z;R_s^2) G(\bm{p};mT).
\end{align}
The yield is written as
\begin{align}
    N_A= \frac{g_A}{2^A} \sum_{\{\pm_i\}_i}\int \biggl[\prod^A_i d^3\bm{r}_i d^3\bm{p}_if_{\pm_i}(\bm{r}_i,\bm{p}_i)\biggr] W_A(\{\bm{r}_i,\bm{p}_i\}^A_{i=1}),
\end{align}
where $\sum_{\{\pm_i\}_i}$ sums over $\{\pm_i\}_{i=1}^A \in \{+,-\}^A$.
The cumulant expansion of the hot-spot phase-space distribution is given by
\begin{align}
  \frac{f_\pm(\bm{z}_i)}{N_p}
  &= \int \frac{d^6\bm\aux_i}{(2\pi)^6}
  e^{-\imag\bm\aux_i\cdot \bm{z}_i} e^{\imag\bm\aux_i\cdot \bm{a}_i - \frac12 \bm{t}_i^\mathrm{T} \mathcal C_2 \bm{t}_i},
\end{align}
where $\bm{a}_i \eqdef \sqrt{2}(0,0,\pm_i aR_s/\sigma_A,0,0,0)^\mathrm{T}$,
and $\mathcal C_2$ is the same as the simple Gaussian case~\eqref{eq:example.gauss.C2}.
The yield is written by the cumulants as
\begin{align}
  N_A
  &= \frac{g_A 8^{A-1} N_p^A}{2^A} \sum_{\{\pm_i\}_i}
    \int \prod_{i=1}^{A-1} \frac{
      d^6\bm\Aux_i e^{-\frac12 \bm\Aux_i^\mathrm{T} (\mathcal C_2 + \mathcal I_6)\bm\Aux_i +\imag \bm\Aux_i\cdot\bm{A}_i}}%
      {(2\pi)^3} \nonumber \\
  &= g_A 8^{A-1} N_p^A \det(\mathcal C_2 + \mathcal I_6)^{-(A-1)/2} \nonumber \\ &\quad \times
    \frac1{2^A} \sum_{\{\pm_i\}_i} e^{-\frac12\sum_{i=1}^{A-1} \bm{A}_i^\mathrm{T} (\mathcal C_2 + \mathcal I_6)^{-1} \bm{A}_i},
\end{align}
where $\bm{A}_i \eqdef \sum_{j=1}^A O_{ij} \bm{a}_j$. The exponent is calculated as
\begin{align}
  & \sum_{i=1}^{A-1} \bm{A}_i^\mathrm{T} (\mathcal C_2 + \mathcal I_6)^{-1} \bm{A}_i \nonumber \\
  &= \sum_{i=1}^{A} \bm{a}_i^\mathrm{T} (\mathcal C_2 + \mathcal I_6)^{-1} \bm{a}_i
    - \bm{A}_A (\mathcal C_2 + \mathcal I_6)^{-1} \bm{A}_A \nonumber \\
  &= \frac 1{2R_s^2/\sigma_A^2 + 1} \biggl[
      \sum_{i=1}^A \bm{a}_{i,\bm{r}}^2
    -\frac1{A} \biggl(\sum_{i=1}^A \bm{a}_{i,\bm{r}}\biggr)^2\biggr] \nonumber \\
  &+ \frac 1{2mT\sigma_A^2 + 1} \biggl[
      \sum_{i=1}^A \bm{a}_{i,\bm{p}}^2
    -\frac1{A} \biggl(\sum_{i=1}^A \bm{a}_{i,\bm{p}}\biggr)^2\biggr] \nonumber \\
  &= \frac{A \Var[\{\bm{a}_{i,\bm{r}}\}_i]}{2R_s^2/\sigma_A^2 + 1}
    +\frac{A \Var[\{\bm{a}_{i,\bm{p}}\}_i]}{2mT\sigma_A^2 + 1},
\end{align}
where $O^TO = \mathrm 1$ has been used to obtain the second line,
$\bm{a}_{i,\bm{r}}$ and $\bm{a}_{i,\bm{p}}$ are the coordinate and momentum parts, respectively, of the vector $\bm{a}_i$,
and $\Var[\cdots]$ is the variance.
For the present case, the momentum part vanishes: $\Var[\{\bm{a}_{i,\bm{p}}\}]=0$.
The coordinate part is
\begin{align}
  \Var[\{\bm{a}_{i,\bm{r}}\}_i]
  &= \frac{2a^2R_s^2}{\sigma_A^2}
    \biggl[1 - \biggl(\frac1A\sum_{i=1}^A \pm_i1\biggr)^2\biggr] \nonumber \\
  &= \frac{2a^2R_s^2}{\sigma_A^2}
    \Bigl[1 - \Bigl(\frac{2r_{\{\pm_i\}} - A}A\Bigr)^2\Bigr],
\end{align}
where $r_{\{\pm_i\}}$ is the number of $+$'s in $\{\pm_i\}_{i=1}^A$.
Finally,
\begin{align}
  & \frac1{2^A} \sum_{\{\pm_i\}_i} e^{-\frac12\sum_{i=1}^{A-1} \bm{A}_i^\mathrm{T} (\mathcal C_2 + \mathcal I_6)^{-1} \bm{A}_i} \nonumber \\
  &= \frac{1}{2^A} \sum_{\{\pm_i\}_i} e^{-\frac{a^2}2\frac{2R_s^2/\sigma_A^2}{2R_s^2/\sigma_A^2 + 1} A[1-(2r_{\{\pm_i\}}/A-1)^2]} \nonumber \\
  &= \frac{1}{2^A} \sum_{r=0}^A \Binomial Ar \alpha^{A[1-(2r/A-1)^2]}.
\end{align}

\section{Derivation of Eq.~\eqref{eq:example.elliptic.flow}: The elliptic flow for the anisotropic distribution function}%
\label{app:example.elliptic.flow}
Here, we see how the analytic result Eq.~\eqref{eq:example.elliptic.flow} is obtained.
Using $\cos2\phi_p = \cos^2\phi_p - \sin^2\phi_p = (p_x^2 - p_y^2)/(p_x^2 + p_y^2)$,
the elliptic-flow coefficient can be calculated by
\begin{align}
  v_2 &= \frac{
    \int d^3\bm{r}d^3\bm{p} f(\bm{r},\bm{p}) \frac{p_x^2 - p_y^2}{p_x^2+p_y^2}
    }{\int d^3\bm{r}d^3\bm{p} f(\bm{r},\bm{p})}.
\end{align}
To perform the integration, we consider the Schwinger parametrization $1/(p_x^2+p_y^2) = \int_0^\infty dt e^{-t(p_x^2+p_y^2)}$:
\begin{align}
  v_2 &= \frac{\int_0^\infty dt \int d^3\bm{r}d^3\bm{p} f(\bm{r},\bm{p}) e^{-t(p_x^2+p_y^2)} (p_x^2 - p_y^2)}%
    {\int d^3\bm{r}d^3\bm{p} f(\bm{r},\bm{p})} \nonumber \\
  &= \frac{\int_0^\infty dt \int d^6\bm{z}' e^{-\frac 12 \bm{z}'^{\mathrm{T}} \mathcal C_2(t)^{-1} \bm{z}'} (p_x^2 - p_y^2)}%
    {\int d^6\bm{z}' e^{-\frac 12 \bm{z}'^{\mathrm{T}} \mathcal C_2(t=0)^{-1} \bm{z}'}},
    \label{eq:example.elliptic.flow.gauss-integration}
\end{align}
where $\bm{z}' \eqdef \tBinomial{\bm{r}}{\bm{p}}$, and
\begin{align}
  \mathcal C_2(t)^{-1} &= \begin{pmatrix}
    \frac{\mathcal I_3 + \frac mT(\mathcal I_\perp +\varepsilon\lambda_3)^2}{R_s^2} &
    -\frac{\mathcal I_\perp +\varepsilon\lambda_3}{TR_s} \\
    -\frac{\mathcal I_\perp +\varepsilon\lambda_3}{TR_s} &
    \frac{1}{mT}\mathcal I_3 + 2t\mathcal I_\perp
  \end{pmatrix}.
\end{align}
The determinant is $\det \mathcal C_2(t) = [\det \mathcal C_2(t)^{-1}]^{-1} = (R_s^2mT)^3/(1+\alpha_+ t)(1+\alpha_- t)$
with $\alpha_\pm \eqdef 2mT[1+ (m/T)(1\pm\varepsilon)^2]$,
and the second-order phase-space cumulants become
\begin{align}
  \mathcal C_2(t) &= [\mathcal C_2(t)^{-1}]^{-1} = \begin{pmatrix}
    R_s^2 \frac{1+2mT \mathcal I_\perp t}{1+\alpha \mathcal I_\perp t} &
    mR_s \frac{\mathcal I_\perp + \varepsilon \lambda_3}{1+\alpha \mathcal I_\perp t} \\
    mR_s \frac{\mathcal I_\perp + \varepsilon \lambda_3}{1+\alpha \mathcal I_\perp t} &
    \frac{\alpha}{2} \frac{1}{1+\alpha \mathcal I_\perp t}
  \end{pmatrix},
\end{align}
where $\alpha \eqdef \diag(\alpha_+, \alpha_-, 2mT)$.
Now, we are ready to evaluate Eq.~\eqref{eq:example.elliptic.flow.gauss-integration}:
\begin{align}
  v_2
  &= \frac{
      \int_0^\infty dt
      \sqrt{\frac{(2\pi)^6 (R_s^2mT)^3}{(1+\alpha_+ t)(1+\alpha_- t)}}
      \frac12 \bigl(\frac{\alpha_+}{1+\alpha_+ t} - \frac{\alpha_-}{1+\alpha_- t}\bigr)
    }{\sqrt{(2\pi)^6 (R_s^2mT)^3}} \nonumber \\
  &= \frac{\alpha_+ - \alpha_-}{2} \int_0^\infty \frac{dt}{(1+\alpha_+ t)^{3/2}(1+\alpha_- t)^{3/2}} \nonumber \\
  &= -\frac{1}{\alpha_+ - \alpha_-}
    \left[ \alpha_+ \sqrt{\frac{1+\alpha_- t}{1+\alpha_+ t}}
      + \alpha_- \sqrt{\frac{1+\alpha_+ t}{1+\alpha_- t}}
    \right]_{t=0}^{t=\infty} \nonumber \\
  &= \frac{\alpha_+ + \alpha_- - 2\sqrt{\alpha_+\alpha_-}}{\alpha_+ - \alpha_-}
  = \frac{(\sqrt{\alpha_+} - \sqrt{\alpha_-})^2}{(\sqrt{\alpha_+})^2 - (\sqrt{\alpha_-})^2} \nonumber \\
  &= \frac{\sqrt{\alpha_+} - \sqrt{\alpha_-}}{\sqrt{\alpha_+} + \sqrt{\alpha_-}}.
\end{align}
Finally, we can eliminate the common factor $2mT$ as $\rho_\pm \eqdef \alpha_\pm / 2mT$
to obtain Eq.~\eqref{eq:example.elliptic.flow}.

\section{Importance sampling of Monte-Carlo integration of light-nuclei yields}
\label{app:importance-sampling}

We here give a method to evaluate the light-nuclei yield
under the blast-wave--type anisotropic flow configuration \eqref{eq:flow-blast3}:
The direct numerical integration for the light-nuclei yield given by the integral \eqref{eq:coale-1}
has unrealistically large computational cost in particular for a larger mass number $A$.
For example, we naively need to perform the 18-dimensional integration for the triton yield ($A=3$).
Instead of the direct numerical integration, we here consider the Monte-Carlo integration.
However, even with the Monte-Carlo integration,
the convergence is not sufficient
within realistic statistics with a larger $A$ due to the curse of dimensionality.
We consider the importance sampling of the Monte-Carlo integration
by utilizing the fact that the phase-space distribution
becomes multivariate Gaussian distribution for the case $u_n=0$.
We sample the phase-space coordinates following the integrand (including the Wigner function and $A$ phase-space distributions)
of the case $u_n=0$ and evaluate the contribution of non-vanishing $u_n$.

The phase-space distribution for $u_n = 0$ is written as
\begin{align}
  f_0(\bm{r},\bm{p}) &= f_z(r_z,p_z)
    e^{-\frac12 (\bm{z}_x^\mathrm{T} A_0 \bm{z}_x + \bm{z}_y^\mathrm{T} A_0 \bm{z}_y)}, \\
  A_0 &:= \begin{pmatrix}
    \frac1{2R^2} (1+\frac mT) \sigma_A^2 & -\frac{1}{2TR} \\
    -\frac{1}{2TR} & \frac1{2Tm\sigma_A^2}
  \end{pmatrix},
\end{align}
where the $(r_z,p_z)$-part is factorized as $f_z(r_z,p_z)$, and
$\bm{z}_{x,y} = (r_{x,y}/\sigma_A, \sigma_A p_{x,y})$.
The phase-space density for the non-vanishing $u_n$ can be expressed by
\begin{align}
  f(\bm{r},\bm{p})
    &= f_0(\bm{r},\bm{p}) e^{-V(\bm{r},\bm{p})}, \\
  V(\bm{r},\bm{p})
    &:= \frac{m}{2T}
      \Bigl[\frac{\bm{r}_\perp^2}{R^2} \delta(\delta+2) - 2\frac{\bm{r}_\perp\cdot\bm{p}_\perp}{mR}\delta\Bigr],
\end{align}
where $\delta(\bm{r}) = 2 u_n\cos(n\phi_s)$,
$\bm{r}_\perp = (r_x, r_y)$, and
$\bm{p}_\perp = (p_x, p_y)$.
With this setup, the $(r_z,p_z)$ integration can be factorized and analytically performed
so that we focus on the integrations by $(r_x,p_x,r_y,p_y)$ hereafter.

For the importance sampling,
we first sample the $A$ phase-space coordinates according
to the multivariate Gaussian distribution given by
$[\prod_{i=1}^A f_0(\bm{r}_i,\bm{p}_i)]W_A(\{\bm{r}_i,\bm{p}_i\}_{i=1}^A)$.
This can be done by expressing the distribution as $\exp[-\frac12\bm{Z}^\mathrm{T} \mathcal{C}^{-1} \bm{Z}]$,
diagonalizing the covariance matrix $\mathcal{C}$ to obtain the eigenmodes (i.e., the principal components) of $\bm{Z}$,
and randomly sampling the eigenmodes according to the Gaussian random numbers with the variance determined by the eigenvalues.
Using the sampled phase-space coordinates, $\{\bm{r}_i^{(1)},\bm{p}_i^{(1)}\}_{i=1}^A, \{\bm{r}_i^{(2)},\bm{p}_i^{(2)}\}_{i=1}^A, \dots$, we evaluate the integration by
\begin{align}
  N_A
    &= g_A \int \biggl[\prod_{i=1}^A d^3\bm{r}_i d^3\bm{p}_i f(\bm{r}_i,\bm{p}_i)\biggr]W_A(\{\bm{r}_i,\bm{p}_i\}_{i=1}^A) \nonumber \\
    &= C \frac1{N_s} \sum_{s=1}^{N_s}
    e^{-\sum_{i=1}^A V(\bm{r}_i^{(s)}, \bm{p}_i^{(s)})},
\end{align}
where $C = g_A N_p^{A} 4^{A-1} \sqrt{\det\mathcal C/(R_s^2mT)^{2A}}[(R_s^2 + \sigma_A^2/2)(mT + 1/2\sigma_A^2)]^{-(A-1)/2}$ is a constant,
$N_s$ is the sample size,
and $\cdots^{(s)}$ is the index of the sample point.

We further tweak the sampling distribution
to avoid the unstable behavior of the Monte-Carlo integration
for large values of $\bm{z}_i$ by mixing the gamma distribution with the normal distribution.
We modify the sampling probability of $n(=4A)$ random variables $\bm{x} = (x_1,\dots,x_n)$
of the standard normal distribution used in the above sampling as
\begin{align}
  & \Pr(\bm{x};\epsilon,\beta) d^n\bm{x} \nonumber \\
  &= \biggl[(1-\epsilon) \frac1{Z_1} e^{-\frac12 \bm{x}^2} + \epsilon \frac 1{Z_2}e^{-|\bm{x}|/\beta}\biggr] d^n\bm{x} \nonumber \\
  &= (1-\epsilon) \frac{e^{-\frac12 \bm{x}^2} d^n\bm{x}}{(2\pi)^{n/2}}
    + \epsilon \frac{r^{n-1} e^{-r/\beta} dr}{\beta^{n-1} \Gamma(n)} \frac{d\Omega_{n-1}}{S_{n-1}}.
\end{align}
where the parameters $\epsilon=1/32$ and $\beta=2$ are
the fraction and scale of the gamma distribution,
$Z_1=(2\pi)^{n/2}$ and $Z_2=S_{n-1} \beta^{n-1} \Gamma(n)$ are
the normalization for $e^{-\frac12 \bm{x}^2}d^n\bm{x}$ and $e^{-|\bm{x}|/\beta}d^n\bm{x}$, respectively,
and $S_{n-1} = 2\pi^{n/2}/\Gamma(n/2)$ is the surface area of $(n-1)$-hypersphere.
In the third line, $r=|\bm{x}|$ and $d\Omega_{n-1}$
are the radius and the infinitesimal element of the angular integration in the $n$-dimensional space.
With this modified distribution, the integration is evaluated as
\begin{align}
  N_A &= C \frac 1{N_s} \sum_{s=1}^{N_s}
    w^{(s)}
    e^{-\sum_{i=1}^A V(\bm{r}_i^{(s)}, \bm{p}_i^{(s)})},
\end{align}
where the reweighting factor reads
\begin{align}
  w^{(s)}
  &= w(\bm{x}^{(s)};\epsilon,\beta) \nonumber \\
  &= \frac{\frac1{(2\pi)^{n/2}} e^{-\frac12\bm{x}^2}}{
    (1-\epsilon) \frac1{(2\pi)^{n/2}} e^{-\frac12\bm{x}^2}
    + \epsilon \frac{1}{S_{n-1}\beta^n\Gamma(n)} e^{-|\bm{x}|/\beta}}.
\end{align}

\bibliographystyle{apsrev4-2}
\bibliography{lightnucl}
\end{document}